\documentclass[usenatbib,useAMS,usegraphicx]{mn2e}

\newcommand{\xmm}{\textit{XMM-Newton}}
\newcommand{\chandra}{\textit{Chandra}}
\newcommand{\asca}{\textit{ASCA}}
\newcommand{\rosat}{\textit{ROSAT}}
\newcommand{\aciss}{\textsc{acis-s}}
\newcommand{\rgs}{\textsc{rgs}}
\newcommand{\mekal}{\textsc{mekal}}
\newcommand{\xspec}{\textsc{xspec}}
\newcommand{\rmf}{\textsc{rmf}}
\newcommand{\arf}{\textsc{arf}}
\newcommand{\fits}{\textsc{fits}}
\newcommand{\phabs}{\texttt{phabs}}
\newcommand{\xmekal}{\texttt{mekal}}
\newcommand{\xvmekal}{\texttt{vmekal}}
\newcommand{\mkcflow}{\texttt{mkcflow}}
\newcommand{\vmcflow}{\texttt{vmcflow}}
\newcommand{\ion}[2]{#1\,\textsc{#2}}
\newcommand{\diff}{\mathrm{d}}
\newcommand{\erf}{\mathop{\rm erf}\nolimits}
\newcommand{\mysub}[2]{\ensuremath{#1_{\mathrm{#2}}}}
\newcommand{\unit}[1]{\ensuremath{\mathrm{\, #1}}}
\newcommand{\unitpow}[2]{\ensuremath{\unit{#1}^{#2}\,}}
\newcommand{\subsun}[1]{\ensuremath{\mathrm{\, #1_{\odot}}}}
\newcommand{\pcmsq}{\unitpow{cm}{-2}}
\newcommand{\Kpcmcu}{\unit{K}\unitpow{cm}{-3}}
\newcommand{\Msun}{\subsun{M}}
\newcommand{\kmps}{\unit{km}\unitpow{s}{-1}}
\newcommand{\kmpspMpc}{\ensuremath{\unit{km}\unit{s}^{-1}\unitpow{Mpc}{-1}}}
\newcommand{\Zsun}{\subsun{Z}}
\newcommand{\omegam}{\ensuremath{\Omega_{\mathrm{M}}}}
\newcommand{\omegal}{\ensuremath{\Omega_{\mathrm{\Lambda}}}}

\begin{document}

\title[Metallicity variations in cooling flows]{
  Some effects of small-scale metallicity variations in cooling flows}
\author[R.G. Morris and A.C. Fabian]{
  R. Glenn Morris\thanks{E-mail: gmorris@ast.cam.ac.uk} and A. C. Fabian\\
  Institute of Astronomy, Madingley Road, Cambridge CB3 0HA}
\maketitle

\begin{abstract}
  In an attempt to reconcile recent spectral data with predictions of the
  standard cooling flow model, it has been suggested that the metals in the
  intracluster medium (ICM) might be distributed inhomogeneously on small
  scales. We investigate the possible consequences of such a situation
  within the framework of the cooling flow scenario. Using the standard
  isobaric cooling flow model, we study the ability of such metallicity
  variations to preferentially suppress low-temperature line emission in
  cooling flow spectra. We then use simple numerical simulations to
  investigate the temporal and spatial evolution of the ICM when the metals
  are distributed in such a fashion. Simulated observations are used to
  study the constraints real data can place on conditions in the ICM\@. The
  difficulty of ruling out abundance variations on small spatial scales
  with current observational limits is emphasized. We find that a bimodal
  distribution of metals may give rise to interesting effects in the
  observed abundance profile, in that apparent abundance gradients with
  central abundance drops and off-centre peaks, similar to those seen
  recently in some clusters, are produced. Different elements behave in
  different fashion as governed by the temperature dependence of their
  equivalent widths. Our overall conclusion is that, whilst this process
  alone seems unlikely to be able to account for the sharp reduction in low
  temperature emission lines seen in current spectral data, a contribution
  at some level is possible and difficult to rule out. The possibility of
  small-scale metallicity variations should be considered when analysing
  high resolution cluster X-ray spectra.
\end{abstract}


\section{Introduction}

Data from the latest generation of X-ray satellites, \chandra{} and \xmm,
are forcing us to re-examine some of the basic tenets of the traditional
cooling flow model \citep[e.g.,][]{fabi94}. Various authors
\citep[e.g.,][]{kaas01,pete01,tamu01} have drawn attention to the
discrepancy between the predictions of the standard cooling flow model and
observed spectra of cluster central regions. The expected emission lines
for several important species (e.g.\ the \ion{Fe}{xvii}{} 15 and 17\,\AA{}
lines) do not seem to be present at the levels which simple models would
expect. This is a trend that appears to be repeated in many clusters that
have traditionally been thought to harbour cooling flows \citep{pete02}.
Lines such as these are important because they are strong indicators of low
($\la 1 \unit{keV}$) temperature gas (see Section~\ref{sec:icm_thermo}). In
a standard cooling flow, in which gas is cooling down to essentially zero,
we expect a significant flux in such lines.

Several ideas \citep[e.g.,][]{fabi01a,pete01} have been put forward to
explain this discrepancy. Here we focus on just one of these, the
suggestion that the intracluster medium (ICM) metals might be distributed
inhomogeneously on small-scales. We examine in detail the consequences of
such a scenario, investigating the effects of an ICM metallicity which
varies on small, unresolved scales (sub-kpc, say). This idea is no more
than a minor extension of the multiphase cooling flow model \citep{nuls86},
which has always relied upon the coexistence of phases with different
densities and temperatures at the same radius in the ICM\@. We merely allow
the chemical composition of these phases to vary as well. This idea is
attractive in the context of cooling flows since it leaves the spectra of
high ($\ga 1\unit{keV}$) temperature gas essentially unchanged, whilst
reducing the line emission from low temperature gas. It (potentially)
enables us to reconcile the data and models whilst making the smallest
conceptual changes to the standard model.

Indeed, given that a suppression of thermal conduction (electron motion)
implies an even more severe reduction in the freedom of heavy ions to
diffuse through the ICM, we might expect small-scale metallicity variations
to be a natural consequence of the multiphase cooling flow model, in which
conduction is by necessity heavily reduced. Conversely, if ICM thermal
conduction is relatively efficient (as has been suggested recently by
several authors, e.g., \citealt{nara01,voig02}) then this does not
necessarily imply that the same is true of ion motion.

In Section~2 we examine the effects of an inhomogeneous metallicity within
the framework of the standard isobaric cooling flow model
\citep[e.g.,][]{john92}, quantifying the reduction in equivalent width that
can be expected for several important ICM species as the degree of
inhomogeneity is varied. In Section~3 we outline an improved numerical
model that allows us to carry out simulations with spatial and temporal
evolution, and to produce simulated observations of \chandra{} spectral
data. Section~4 describes some of the results of this model for a simple
case of metallicity inhomogeneities. We find that small-scale abundance
variations may give rise to the appearance of abundance gradients where
none in fact exist, with the results for individual species being
controlled by the temperature dependence of their equivalent widths. We
stress the difficulty of resolving such variations without a combination of
good spatial and spectral resolution. In the final Section we make some
discussion concerning the likelihood of such small-scale metallicity
variations in the ICM\@. Given that direct detection of such structures is
currently unfeasible, we put forward some suggestive circumstantial
evidence.

Throughout this paper we assume: $H_{0} = 50$\kmpspMpc; $\omegam = 1.0$,
$\omegal = 0.0$. We make use of \xspec{} version 11.

\section{Cooling Flow Spectra}\label{sec:cf_spec}

\subsection{ICM thermometers}\label{sec:icm_thermo}

\begin{figure}
  \begin{center}
    \includegraphics[draft=false,angle=0,width=1.0\columnwidth]
    {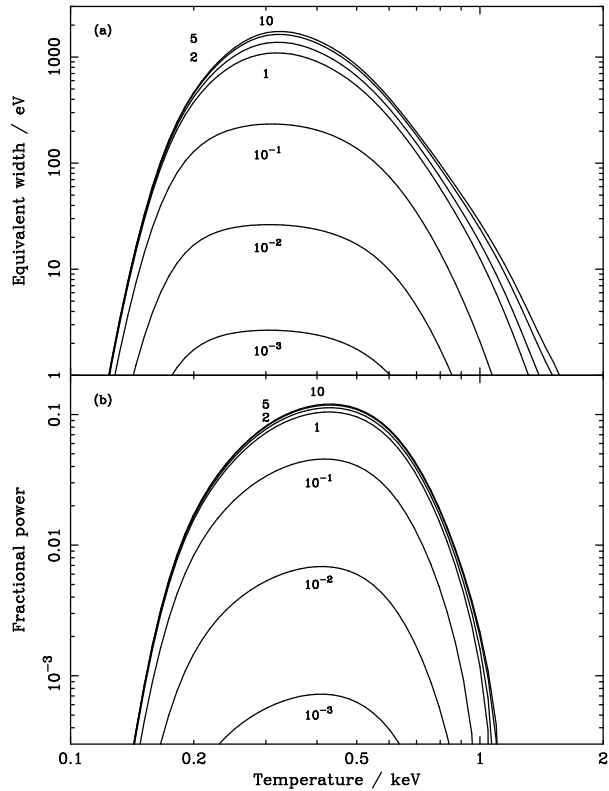}
  \end{center}
  \caption{
    Behaviour of the \ion{Fe}{xvii}{} 15.0\,\AA{} line in a coronal
    equilibrium single temperature plasma. The curve labels refer to
    metallicity in solar units. The top panel shows the equivalent width in
    eV, the bottom panel shows the fractional power (in the X-ray waveband
    0.1--10.0\unit{keV}) contained in this line.
  }
  \label{fig:15ang}
\end{figure}

Fig.~\ref{fig:15ang}a shows the temperature dependence of the equivalent
width of the 15\,\AA{} \ion{Fe}{xvii}{} line, for various metallicities,
for a single temperature plasma, as calculated using the \mekal{} plasma
code \citep{mewe95,kaas93}. This code computes the bremsstrahlung continuum
and line radiation from a plasma (of up to 15 elements) in the coronal
equilibrium approximation \citep[e.g.,][]{bric95} over the temperature
range $10^{4}$--$10^{9}$\unit{K}. The equivalent width curves are sharply
peaked around 0.3 keV ($\sim 10^{6}$\unit{K}), so these emission lines are
sensitive indicators of gas at such temperatures. Furthermore, they are
extremely strong -- Fig.~\ref{fig:15ang}b shows the fraction of the power
(in the X-ray waveband 0.1--10.0\unit{keV}) being emitted at any
temperature in this iron line. For a solar metallicity plasma, it peaks
around the 10 per cent level. The results are similar for the 17\,\AA{}
line. Consequently, these lines are highly important coolants below
1\unit{keV}, and therefore in the standard cooling flow model (in which gas
cools from a temperature of several \unit{keV} to essentially zero), we
expect a strong flux in these lines. The actual fluxes seen in many cooling
flow clusters are much weaker than would be expected
\citep[e.g.,][]{kaas01,pete01,tamu01}.

\subsection{Bimodal metallicities}

\subsubsection{Fixed temperature plasmas}

Several ideas \citep[e.g.,][]{fabi01a,pete01} have been put forward to
explain this missing low temperature flux. Here we focus on just one of
these, the suggestion that the ICM metals might be inhomogeneously
distributed on small-scales. The essence of this idea is as follows. At any
fixed single temperature, a two-component plasma consisting of metal-rich
and metal-poor `phases' (note that the usage of the term phase here is not
necessarily strictly the same as in traditional multiphase cooling flow
models) is spectroscopically indistinguishable from a homogeneous plasma of
some mean metallicity $\bar{Z} = \mysub{f}{hi} \mysub{Z}{hi} + (1 -
\mysub{f}{hi}) \mysub{Z}{lo}$, where the metal-rich phase has a metallicity
\mysub{Z}{hi} and accounts for a mass fraction \mysub{f}{hi}, and the
metal-poor phase has a metallicity \mysub{Z}{lo}. For simplicity (and
without loss of generality), we take $\mysub{Z}{lo} \equiv 0$.
$\mysub{f}{hi} = 1$ corresponds to a uniform plasma, and decreasing values
of \mysub{f}{hi} to increasingly segregated plasmas. Throughout this paper,
we leave the He abundance fixed at the solar value, allowing the heavy
elements to vary as specified by $Z$. We use the solar abundance ratios of
\citet{ande89}, hereinafter \Zsun.

The indistinguishability of the spectra in these two cases is a simple
consequence of the fact that the strength of the continuum radiation is
independent of $Z$, whilst that of the emission lines is directly
proportional to $Z$. The reduction in the mass fraction of the enriched
phase is therefore offset by the increased line strength. Alternatively, in
the coronal limit, the local concentration of the heavy elements makes no
difference to their radiation: a given number of heavy ions will radiate in
the same way whether uniformly dispersed throughout the emitting volume or
concentrated in a specific region. As an aside, we note that this
simplistic treatment breaks down at very high metallicities ($Z \ga
70\Zsun$) when the free electrons contributed by the metal ions become
significant. This would not invalidate the idea that a mixture of a
metal-rich and a metal-poor plasma can look identical to a uniform
metallicity plasma, it would merely change the dependence of the
normalization of each component on $Z$ from a simple linear relationship to
something more complicated. Such ultra-high metallicities are of no
relevance here though.

\subsubsection{Cooling plasmas}\label{sec:cflow_ew}

The situation becomes more complex when one allows cooling to occur. We
briefly review the derivation of the standard isobaric cooling flow model
\citep[e.g.,][]{john92}. The energy released per unit mass of gas on cooling by
$\diff\,T$ is
\begin{eqnarray*}
\diff\,q & = & \diff\,u + p \diff\,\left(\frac{1}{\rho}\right) =
\diff\left( u + \frac{p}{\rho} \right)
\qquad(\mathrm{constant}\;p)\\
& = & \diff\left( \frac{3}{2} \frac{\mysub{k}{B} T}{\mu \mysub{m}{H}} +
  \frac{n \mysub{k}{B} T}{\rho} \right)
\qquad(\mathrm{ideal}\;\mathrm{gas}).
\end{eqnarray*}

For a constant mass-flow rate $\dot{M}$, the power release is thus
\begin{equation}
  \diff\,L = \frac{5}{2} \frac{\mysub{k}{B}}{\mu\mysub{m}{H}} \dot{M} \diff\,T.
\end{equation}

We also have, from the definition of the cooling function $\Lambda$,
\begin{eqnarray}\nonumber
  \diff\,L & = & \mysub{n}{e} \mysub{n}{H} \Lambda(T,Z) \diff\,V,\\
  \diff\,L_{\nu} & = & \mysub{n}{e} \mysub{n}{H} \Lambda_{\nu}(T,Z) \diff\,V.
\end{eqnarray}

Hence we obtain the spectral power for a steady-state flow cooling from
\mysub{T}{max} to \mysub{T}{min}
\begin{equation}\label{eqn:cflow}
  L_{\nu} = \frac{5}{2} \frac{\mysub{k}{B}}{\mu \mysub{m}{H}} \dot{M}
  \int_{\mysub{T}{min}}^{\mysub{T}{max}}
  \frac{\Lambda_{\nu}(T,Z)}{\Lambda(T,Z)} \diff\,T.
\end{equation}

Thus, we have the simple result that the emission measure of each
temperature component in the flow is inversely proportional to the cooling
function $\Lambda(T)$ at that temperature. This model is implemented by the
\xspec{} package as the \mkcflow{} and \vmcflow{} models\footnote{We draw
attention to the fact that in the \xspec{} \mkcflow{} implementation,
altering the plasma bulk metallicity varies the helium abundance (for
\xspec{} versions prior to 11.1.0ab). This can have a non-negligible effect
on the calculation of equivalent widths when high metallicities are
involved. As a workaround, the \vmcflow{} model may be used to specify
abundances on an element-by-element basis.}. We have constructed an
independent implementation of Eqn.~\ref{eqn:cflow}, calculating cooling
functions by integrating \mekal{} spectra against energy over the
5\unit{eV}--200\unit{keV} range.

\begin{figure}
  \begin{center}
    \includegraphics[draft=false,angle=0,width=1.0\columnwidth]
    {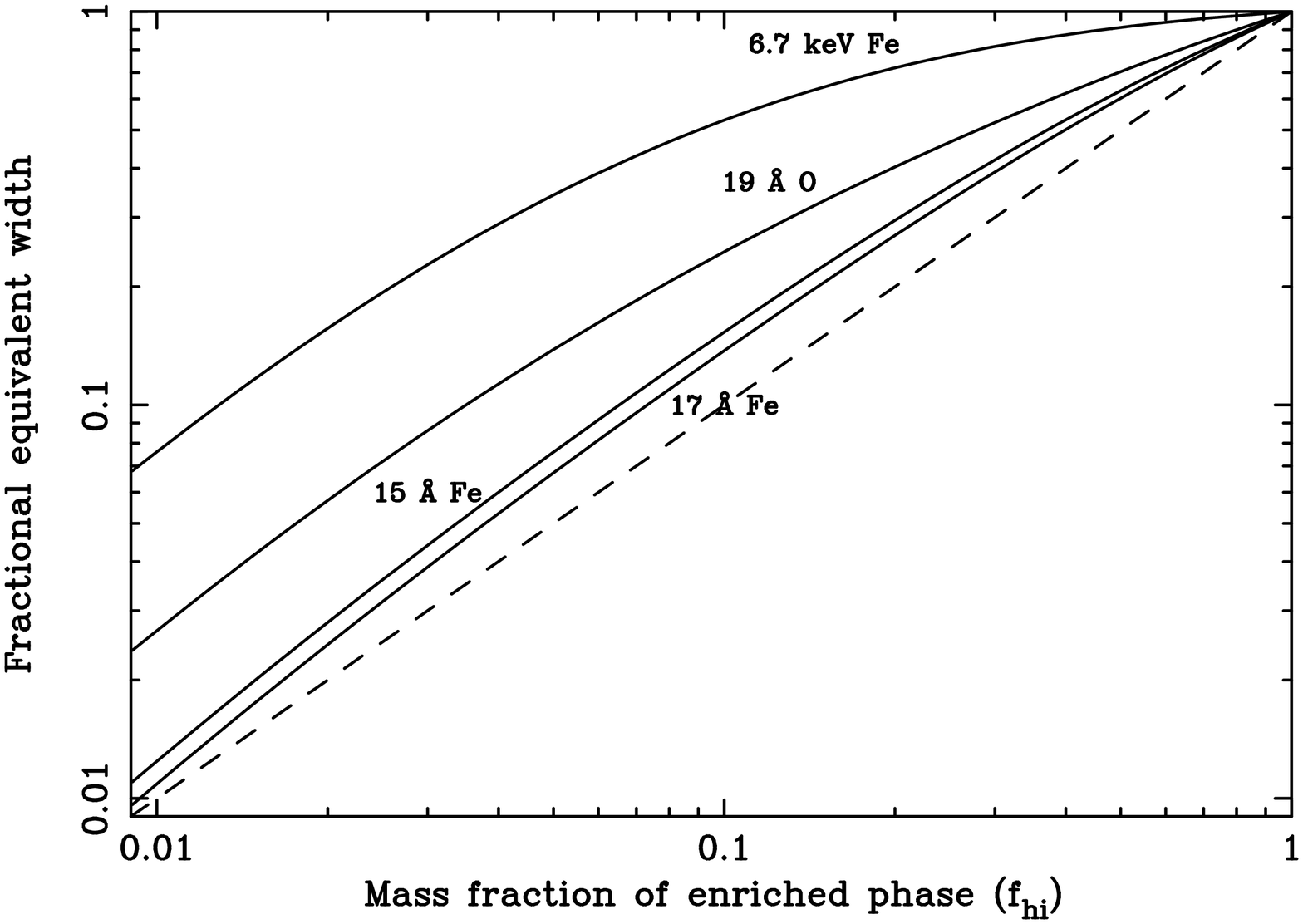}
  \end{center}
  \caption{
    Fractional change in the equivalent width of some important emission
    lines from the cooling flow model of Eqn.~\ref{eqn:cflow}, for
    two-component plasmas, as the mass fraction of the enriched phase is
    varied. The dashed line is the reference line $y = x$. Parameters of
    the cooling flow model: $\bar{Z} = 0.3$\Zsun, $\mysub{Z}{lo} \equiv 0$,
    $\mysub{Z}{hi} = \mysub{f}{hi}^{-1} \bar{Z}$. The integration over
    temperature was performed at 100 points, logarithmically spaced between
    0.0808--8.620\unit{keV}.
  }
  \label{fig:ew_fhi}
\end{figure}

The heating solutions to the cooling flow spectral problem (i.e.\ the
absence of spectral lines characteristic of cool, $\la 1\unit{keV}$, gas)
suppress the low-temperature line emission by raising \mysub{T}{min} to
$\sim 1\unit{keV}$. The bimodal metallicity hypothesis suppresses the
equivalent width (EW) of the low-temperature lines by a different means. In
Fig.~\ref{fig:ew_fhi} we illustrate how the EW of various important
spectral features changes as a function of \mysub{f}{hi} (or
\mysub{Z}{hi}), for the case $\bar{Z} = 0.3\Zsun$. The dashed line shows
what might be the naive expectation, namely that
$\mathrm{EW}(\mysub{f}{hi}) = \mysub{f}{hi} \;\mathrm{EW}(1)$. In no case
is this line followed -- the reduction in the EW is always less than this.
Thus, for example, at $\mysub{f}{hi} = 0.1$, the EW of the Fe K$\alpha$
line is reduced by a factor of about 0.5, that of the 19\,\AA{} O line by
about 0.25, and that of the 15\,\AA{} Fe line by around 0.15. We can
interpret these results as follows.

Recall the definition of equivalent width
\begin{equation}
  \mathrm{EW} \equiv \int_{E_{0}}^{E_{1}} \frac{\mysub{I}{L+C} - \mysub{I}{C}}
  {\mysub{I}{C}}\,\diff E,
\end{equation}
where \mysub{I}{L+C} is the total (line + continuum) intensity at a given
energy, and \mysub{I}{C} is the continuum intensity at the same point.

For fixed, narrow line profiles, we can neglect the integral over energy
and simply compare intensity ratios at the line energy. Making use of the
cooling flow equation Eqn.~\ref{eqn:cflow}, we may write the equivalent
width, EW(H), for a given spectral line (with a narrow energy profile
centred on some energy $E$) from a homogeneous metallicity cooling flow as
\begin{equation}\label{eqn:ew_hom}
  \mathrm{EW(H)} = \frac{\displaystyle\int_{\mysub{T}{min}}^{\mysub{T}{max}}
    \frac{\mysub{\Lambda}{L}(T,E,\bar{Z})}{\Lambda(T,\bar{Z})}
    \diff\,T}
  {\displaystyle\int_{\mysub{T}{min}}^{\mysub{T}{max}}
    \frac{\mysub{\Lambda}{C}(T,E,\bar{Z})}{\Lambda(T,\bar{Z})}
    \diff\,T}.
\end{equation}
In this expression, $\mysub{\Lambda}{L} \mysub{n}{e}\mysub{n}{H}$ would
be the power per unit volume per unit energy radiated by the line component
at a given energy, with \mysub{\Lambda}{C} having an analogous meaning for
the continuum component. Integrating the sum of these two functions over
energy leads to the cooling function $\Lambda(T,Z)$.

For a two-component (`bimodal') cooling flow, in which one component has a
metallicity \mysub{Z}{hi} and accounts for a fraction \mysub{f}{hi} of the
total mass flow rate, and the other component has zero metals, we have the
following expression for the equivalent width, EW(B), of a narrow spectral
line centred at energy $E$:
\begin{eqnarray}\label{eqn:ew_bim}
  \lefteqn{\mathrm{EW(B)}} & & \quad = \mysub{f}{hi}
  \int_{\mysub{T}{min}}^{\mysub{T}{max}}
  \frac{\mysub{\Lambda}{L}(T,E,\mysub{Z}{hi})}{\Lambda(T,\mysub{Z}{hi})}
  \diff\,T \quad /\\\nonumber
  & & \left\{
    \int_{\mysub{T}{min}}^{\mysub{T}{max}}
    \left[\mysub{f}{hi}
      \frac{\mysub{\Lambda}{C}(T,E,\mysub{Z}{hi})}{\Lambda(T,\mysub{Z}{hi})}
      + (1 - \mysub{f}{hi})
      \frac{\mysub{\Lambda}{C}(T,E,0)}{\Lambda(T,0)}
    \right]
    \diff\,T\right\}.
\end{eqnarray}

Our desire is to suppress the equivalent width of the low-temperature lines
in the bimodal case -- that is, the ratio of Eqn.~\ref{eqn:ew_bim} to
Eqn.~\ref{eqn:ew_hom} should be $< 1$ in such cases. There exists no simple
algebraic solution to this form, owing to the non-analytic behaviour of the
$\Lambda$ functions.

Restricting our attention to an emission line that exists only over a
narrow temperature range (for example, the \ion{Fe}{xvii}{} 15\,\AA{}
line, as illustrated in Fig.~\ref{fig:15ang}), however, we may make the
simplification $\int \frac{\mysub{\Lambda}{L}(T)}{\Lambda(T)} \diff\,T
\rightarrow \frac{\mysub{\Lambda}{L}(\bar{T})}{\Lambda(\bar{T})} \delta T$.
In addition, we have the result that
$\mysub{\Lambda}{L}\mysub{n}{e}\mysub{n}{H} \propto
\mysub{n}{e}\mysub{n}{ion} \Rightarrow \mysub{\Lambda}{L} \propto
\mysub{n}{ion} \propto Z$ (at fixed \mysub{n}{H}). Consequently
\begin{equation}
  \frac{\mysub{\Lambda}{L}(\bar{T},E,\mysub{Z}{hi})}
  {\mysub{\Lambda}{L}(\bar{T},E,\bar{Z})} =
  \frac{\mysub{Z}{hi}}{\bar{Z}} = \frac{1}{\mysub{f}{hi}}.
\end{equation}
The ratio of the equivalent widths may therefore be somewhat simplified to
\begin{eqnarray}
  \lefteqn{\frac{\mathrm{EW(B)}}{\mathrm{EW(H)}}} & & \quad\; =
  \frac{\Lambda(\bar{T},\bar{Z})}{\Lambda(\bar{T},\mysub{Z}{hi})}
  \int_{\mysub{T}{min}}^{\mysub{T}{max}}
  \frac{\mysub{\Lambda}{C}(T,E,\bar{Z})}{\Lambda(T,\bar{Z})}
  \diff\,T \quad/\\\nonumber
  & & \left\{
    \int_{\mysub{T}{min}}^{\mysub{T}{max}}
    \left[\mysub{f}{hi}
      \frac{\mysub{\Lambda}{C}(T,E,\mysub{Z}{hi})}{\Lambda(T,\mysub{Z}{hi})}
      + (1 - \mysub{f}{hi})
      \frac{\mysub{\Lambda}{C}(T,E,0)}{\Lambda(T,0)}
    \right]
    \diff\,T\right\}.
\end{eqnarray}
The term involving the ratio of integrals is the ratio of the continuum in
the homogeneous case to that in the bimodal case. By inspection of cooling
flow spectra, this ratio is very close to one. Most of the continuum flux
is contributed by gas at temperatures $\ga 1\unit{keV}$, where the metals
have relatively little influence on the cooling function. Bimodal plasmas
tend to have fractionally stronger continua at low ($\la 1\unit{keV}$)
energies, where low-temperature gas makes a slight contribution. The
reduced cooling function of the low-metallicity phase at low temperatures
enhances its emission measure compared to the mean metallicity case.

In the idealized case of a narrow emission line, the ratio of the
equivalent widths for the two cases therefore reduces to the ratio of the
cooling functions $\Lambda(\bar{T},\bar{Z}) /
\Lambda(\bar{T},\mysub{Z}{hi})$. For a high-temperature ($\ga 1\unit{keV}$)
line, where cooling is essentially independent of metallicity, the ratio of
the equivalent widths reduces to one. For a low-temperature line, then if
the metals were to completely dominate the cooling, we expect the ratio to
reduce to $\bar{Z} / \mysub{Z}{hi} = \mysub{f}{hi}$.

Qualitatively, on increasing the metallicity by a factor $\mysub{f}{hi}^{-1}$,
the emission line strength (at all temperatures) increases proportionately.
Reducing the mass fraction compensates, so that spectra at any fixed
temperature are unchanged. When steady-state cooling is allowed to take
place, however, the faster cooling of the metal-rich gas at low
temperatures reduces its emission measure and suppresses its spectral
contribution.

\begin{figure}
  \begin{center}
    \includegraphics[draft=false,angle=0,width=1.0\columnwidth]
    {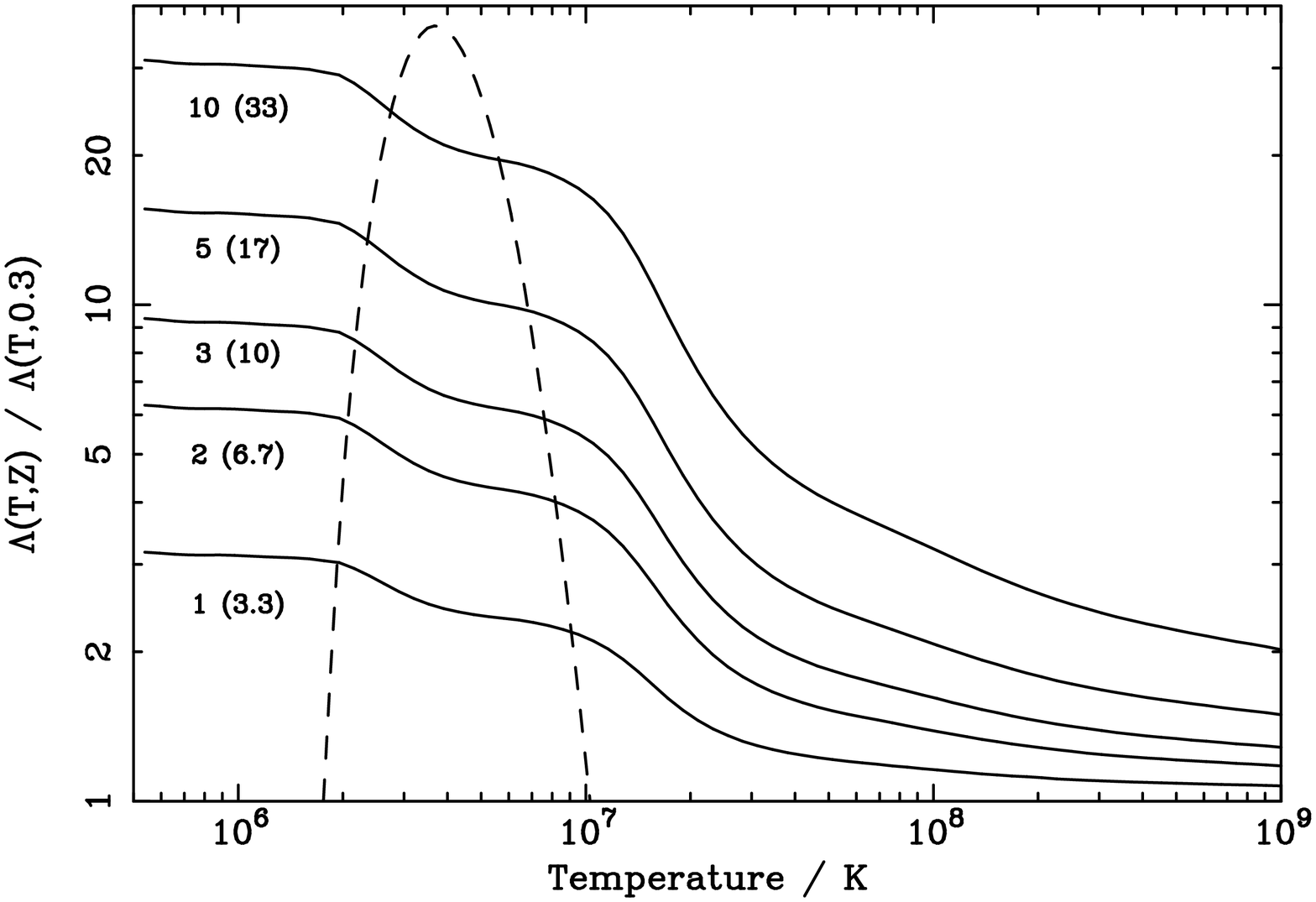}
  \end{center}
  \caption{
    Dependence of the cooling function $\Lambda(T)$ on metallicity. The
    labelled curves show the cooling function for a given metallicity (in
    solar units) relative to that for 0.3\Zsun. The numbers in parentheses
    are the metallicities as multiples of 0.3\Zsun. At low temperatures,
    the curves asymptote out to these values. Above $10^{8}\unit{K}$ the
    cooling is dominated by the H, He continuum; as the temperature
    decreases to $\sim 10^{6}\unit{K}$ Fe becomes increasingly dominant;
    and towards the end of the range O is the major coolant. The dashed
    curve shows the equivalent width of the \ion{Fe}{xvii}{} 15\,\AA{}
    line in arbitrary units.
  }
  \label{fig:lambda_Z}
\end{figure}

The cooling function ratio is plotted for some representative metallicities
in Fig.~\ref{fig:lambda_Z}. Note that metals do not fully dominate the
cooling function until $T \la 10^{6} \unit{K}$. In the temperature range
where the \ion{Fe}{xvii}{} 15\,\AA{} line is strong (overlaid), the ratio of
the cooling function for $Z = 3.0\Zsun$ to that for $Z = 0.3\Zsun$ is in
the range 6--7. This is consistent with the result from
Fig.~\ref{fig:ew_fhi} for this line with $\mysub{f}{hi} = 0.1$, where a
fractional suppression of the EW of order 0.15 is found. This is to be
contrasted with the simple expectation of 0.1. The actual suppression of
this line is not so strong.

In practice for lines existing at intermediate temperatures, or over a
non-negligible range of temperatures, we expect behaviour between these
limits. The iron K line, for example, deriving as it does from gas over a
relatively wide, high temperature range (see Fig.~\ref{fig:eqwidths}), has
its equivalent width suppressed by about a factor of 0.5, for
$\mysub{f}{hi} = 0.1$. If this were the case, then analyses not taking this
into account would have underestimated the mean ICM metallicity. As another
consequence of this, the \textit{relative} suppression of the
\ion{Fe}{xvii}{} line (say) compared to that of the Fe K line, is only a
factor of 0.3. In other words, given a spectrum of unknown metallicity, if
one were to fit an abundance from the iron K line, then attempt to
reproduce the observed \ion{Fe}{xvii}{} line with this abundance, the
observed line would be about a factor of three smaller than predicted. This
is of course appreciably less than the simple `factor of ten' reduction
that might have been expected. Consequently, it seems unlikely that
small-scale metallicity variations alone can be responsible for the very
strong suppression of the equivalent widths of low-temperature emission
lines seen in cooling flow spectra, unless the range of the variations were
to be pushed to extreme levels.

$\mysub{f}{hi} = 0.1$ corresponds to $\mysub{f}{hi} \approx 3\Zsun$. This
may seem like an excessively high metallicity, but consider that supernovae
can be viewed as sources of essentially infinite metallicity. The
\chandra{} observation \citep{iwas01} of 4C55.15 reveals a metallicity
$\approx 2\Zsun$ within the central 50\unit{kpc}, as well as a metal-rich
`plume' extending over $\sim 25\unit{kpc}$ with $Z \approx 8\Zsun$.
Whilst this is clearly an extreme case, if such high-metallicity features
can be produced on these scales, there is no reason they cannot occur on
the much smaller scales we consider.

\section{An Improved Cooling Flow Model}

The standard isobaric cooling-flow model represented by
Eqn.~\ref{eqn:cflow} is a useful tool, but it suffers from a number of
limitations. It assumes a constant pressure, and is inherently
single-phase. It assumes a steady-state with no time dependence.
Furthermore, there is no spatial information, and no attempt to treat the
gravitational potential of the cluster. It is a good first approximation to
the global properties of a cooling flow, but in the new era of high spatial
resolution X-ray spectroscopy and imaging, being able to study the temporal
and spatial evolution of the ICM becomes more and more desirable. An
improved model may be obtained by following the numerical prescription of
\citet{thom88}. The underlying theoretical basis is that of \citet{nuls86}.
We assume spherical symmetry, so that the system can be described by a
one-dimensional model.

\subsection{The dark matter}

We simulate a cluster simply as a two-component system, comprised of hot
gas in the potential well of a dark matter (DM) halo. We represent the
latter by a standard Navarro, Frenk and White (NFW) profile \citep{nava97}.
In order to specify the parameters of the profile, we make use of the
calibrated virial scaling relations of \citet{evra96}:
\begin{eqnarray}
  M(r_{200}) & = & 2.9 \times 10^{15} h_{50}^{-1}
  \left(
    \frac{T}{10\unit{keV}}
  \right)^{\frac{3}{2}} \Msun, \nonumber\\
  r_{200}(T) & = & 3.7 h_{50}^{-1}
  \left(
    \frac{T}{10\unit{keV}}
  \right)^{\frac{1}{2}} \unit{Mpc};\label{eqn:evra}
\end{eqnarray}
where $r_{200}$ is the radius at which the density contrast (i.e.\ density
in units of the critical density) is equal to 200, and serves to define a
virial radius for the cluster \citep[e.g.,][]{cole96}. We complete the
specification of the DM profile by taking a typical cluster value of 5
\citep[e.g.,][]{bull01} for the concentration parameter (i.e.\ the virial
radius in units of the NFW scale radius).

\subsection{The gas}

The gas is modelled as a single-phase medium. The traditional assumption
for cooling flows has been that the unstable nature of the cooling process
would lead to a highly multiphase medium \citep[e.g.,][]{alle00}, i.e.\
phases of different densities and temperatures coexist at the same radius.
Recent observations cast doubt on this assumption, though -- see for
example \citet{mole01}.

For initial conditions, we assume a continuous smooth pressure profile of
the isothermal form, normalized in terms of an (initially arbitrary) outer
boundary pressure at the virial radius. The gas is assigned a density
profile which is proportional to the pressure profile. The scaling is such
that the temperature of the gas at the outer boundary is equal to the
virial temperature of the cluster. Note that when we allow the metallicity
of the gas to vary spatially, the molecular weight $\mu$ is not constant,
and we calculate the outer boundary temperature using a `mean' $\mu$. Also,
in such cases the gas will not be isothermal, since for a smooth pressure
profile to exist, regions of high $\mu$ must be at higher temperatures.

The total mass of gas is calculated via numerical integration of the
density profile.
We adjust the outer boundary pressure so that the system gas mass
corresponds to some appropriate fraction \mysub{f}{gas} of the total
(gas plus DM) mass.
A temperature profile $T(r)$ may then be assigned using the (ideal gas)
equation of state.

\subsection{Evolution of the system}

The system is allowed to evolve under the influence of gravity (neglecting
the gas self-gravity) and radiative energy loss. We simulate the cluster
over the radial range $0 < r < r_{200}$, i.e.\ out to the virial radius.
For boundary conditions, we impose a stationary inner boundary and a
constant pressure outer boundary. The latter is acceptable because
unphysically long time-scales would be required for the effects of cooling
to propagate out as far as the virial radius. The governing differential
equations are represented by the Lagrangian difference scheme of
\citet{thom88}. Put simply, the radial range is divided into a large number
of zones, which are then evolved forward in discrete time-steps. We use an
adaptive step-size that is limited by stability considerations, but also by
the cooling time of the gas, which in the standard isobaric assumption is
given by
\begin{equation}\label{eqn:tcool}
  \mysub{t}{cool} = \frac{\gamma}{\gamma -1}
  \left(
    \frac{\mysub{k}{B}}{\mu \mysub{X}{H}}
  \right)^{2}
  \frac{1}{\mysub{e}{r} p}
  \int_{0}^{T} \frac{T \,\diff T}{\Lambda(T)},
\end{equation}
where: $\gamma$ is the adiabatic index of the plasma; $\mysub{X}{H} =
\mysub{n}{H} \mysub{m}{H} / \rho$ is the hydrogen mass fraction;
$\mysub{e}{r} = \mysub{n}{e} / \mysub{n}{H}$ is the electron density
relative to the hydrogen density. With the cooling function $\Lambda(T)
\sim T^{\frac{1}{2}}$ (at high temperatures) for bremsstrahlung, then at a
fixed pressure $\mysub{t}{cool} \sim T^{\frac{3}{2}}$; whereas for
virially-scaled clusters with $p \propto T$ at a fixed overdensity,
$\mysub{t}{cool} \sim T^{\frac{1}{2}}$.

Since we are concerned only with modelling the gas in the X-ray regime, we
do not follow the cooling below $10^{5}$\unit{K}. At this temperature, the
gas is no longer radiating appreciably in the X-ray waveband (see
Fig.~\ref{fig:lambda}), and may therefore (for our purposes) be ignored
from further consideration. In addition, the cooling time (see
Fig.~\ref{fig:tcool}) is becoming prohibitively short for continued
computation.

If at the end of an iteration step any zone has cooled below
$10^{5}$\unit{K}, it is removed from subsequent analysis, and the remaining
zones are adiabatically expanded to fill the vacated volume. Zones closest
to the loss-site undergo the greatest degree of expansion. Conceptually,
this is intended as a (very) crude representation of the rapid collapse of
cold gas into condensed objects of negligible volume (neglecting any kind
of energy injection at this point from star formation say).

\subsection{The plasma radiation}

We model the radiation coming from the plasma with the \mekal{} spectral
code. Thus, the abundance of the plasma becomes a relevant parameter, and
each radial zone may have a different set of abundances. At the start of a
simulation, spectra for each abundance set are integrated in order to
obtain evaluations of the cooling function $\Lambda(T)$, and the cooling
time integral $\int_{\mysub{T}{min}}^{T} \frac{T \,\diff T}{\Lambda(T)}$ at
discrete temperatures over the range of interest. The relative electron
density is also calculated as a function of temperature. During the course
of a simulation we interpolate amongst the tabulated values of these
functions to find the value appropriate for any particular plasma
temperature.

\begin{figure}
  \begin{center}
    \includegraphics[draft=false,angle=0,width=0.9\columnwidth]
    {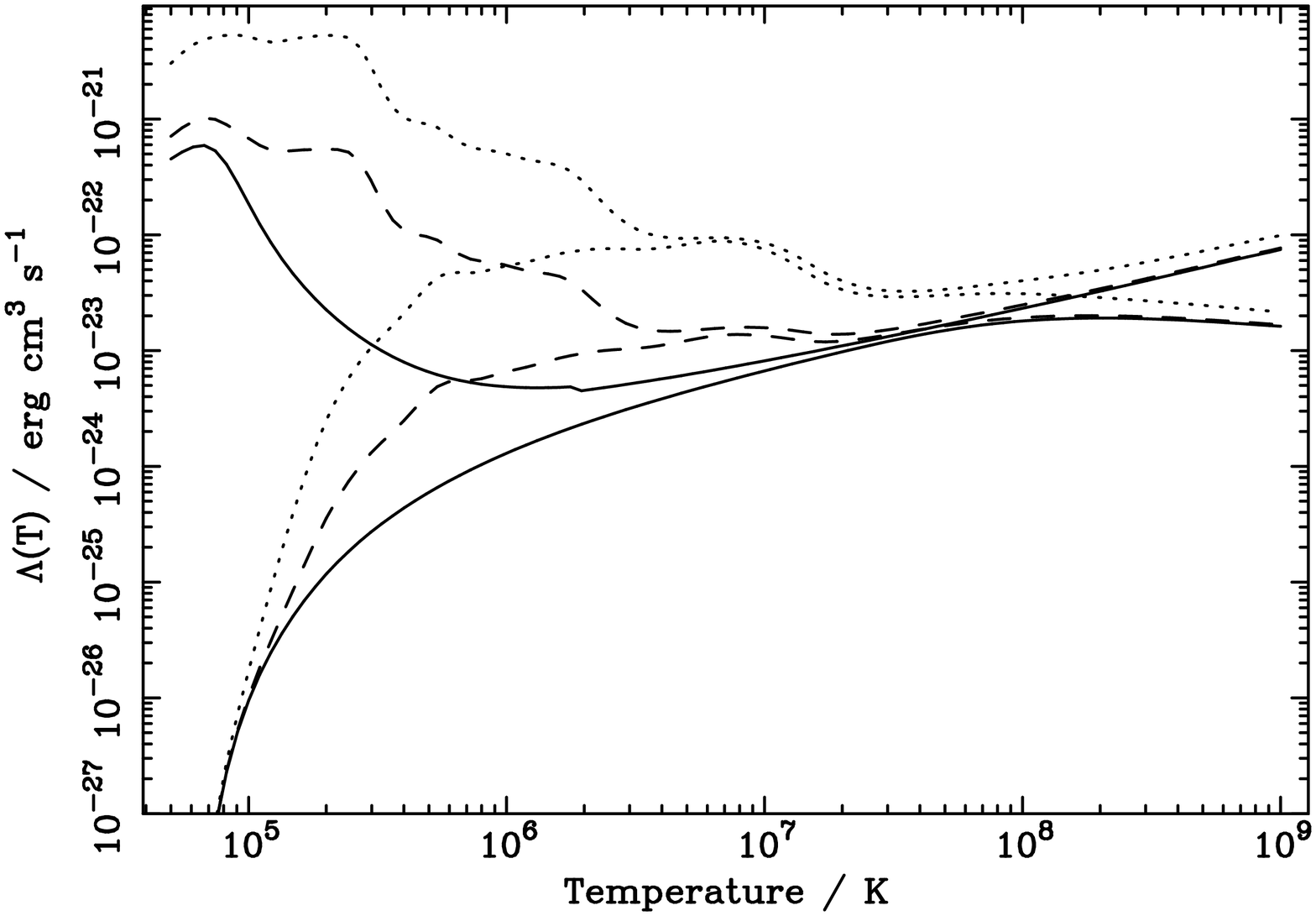}
  \end{center}
  \caption{
    Cooling functions of plasmas of varying metallicities. He is at solar
    abundances in all cases. Solid -- H, He only; dashed -- metals at 0.3
    solar abundance; dotted -- metals at 3.0 solar abundance. The upper
    curve in each case is for all thermodynamically relevant radiation
    (5\unit{eV}--200\unit{keV}); the lower curve is for the X-ray wave-band
    0.1--10\unit{keV}.}
  \label{fig:lambda}
\end{figure}

\begin{figure}
  \begin{center}
    \includegraphics[draft=false,angle=0,width=0.9\columnwidth]
    {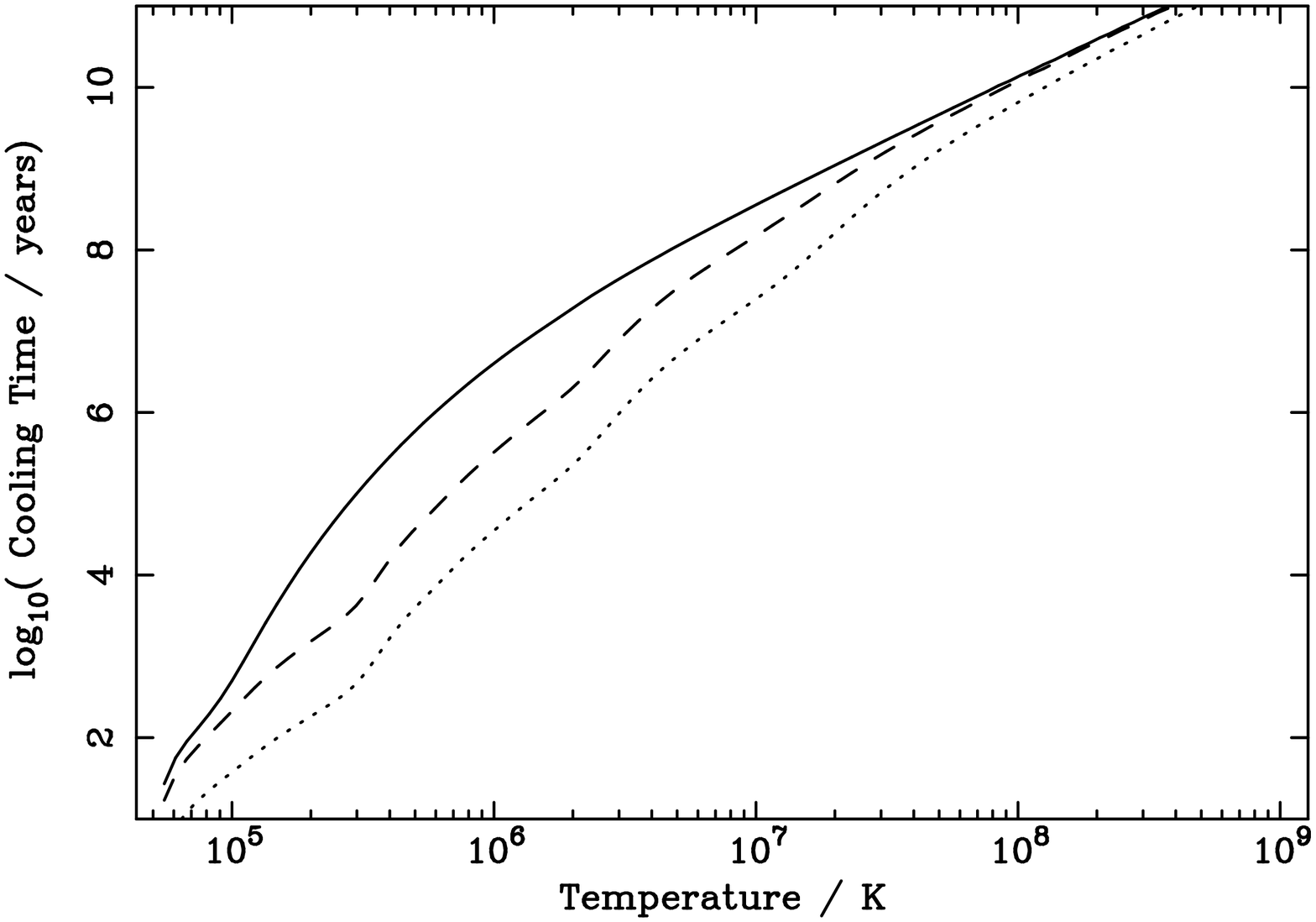}
  \end{center}
  \caption{
    Isobaric cooling times for the plasmas of Fig.~\ref{fig:lambda},
    calculated by integrating the total radiation curves for each abundance
    set. The pressure is $10^6$\Kpcmcu, with the times inversely
    proportional to pressure.
  }
  \label{fig:tcool}
\end{figure}

Besides these essential quantities, it is clearly of great interest to use
\mekal{} to produce synthetic spectra for the plasma. These calculations
need not be performed at every iteration, but only when the physical state
of the system is to be examined. Note that the energy range of interest is
different in the two cases. In the case of $\Lambda(T)$, we are concerned
with all the radiation that acts as a significant energy loss for the
plasma over the temperature range of interest. In practice we use the
energy range 5\unit{eV}--200\unit{keV}. In the case of spectra, only the
much more limited wave-band of, say, 0.1--10\unit{keV}, is of relevance for
comparison with the X-ray observations of \chandra{} and \xmm. These two
functions are displayed in Fig.~\ref{fig:lambda}. Rather than being
interpolated, spectra are calculated directly for each zone.

To facilitate comparison with observations we first integrate along a
line-of-sight to obtain the `spectral surface brightness' at projected
distance $b$ from the centre
\begin{equation}
  S_{\nu}(b) = \int_{b}^{\infty}
  \frac{\epsilon_{\nu}(r) 2 r \,\diff r}{\sqrt{r^{2} - b^{2}}}.
\end{equation}
We then integrate again over an annulus to obtain the spectrum due to the
(distance-weighted) contributions of all the gas along the lines-of-sight
between projected radii $b_{1}$ and $b_{2}$
\begin{eqnarray}
  P_{\nu}(b_{1} \leq b \leq b_{2}) & = &
  \int_{b_{1}}^{b_{2}}\diff b\; 2 \upi b
  \int_{b}^{\infty}
  \frac{\epsilon_{\nu}(r) 2 r \,\diff r}
  {\sqrt{r^{2} - b^{2}}}\\
  & = & 4 \upi [ F(b_{1}) - F(b_{2}) ],\qquad \mathrm{where}\nonumber\\
  F(b) & \equiv & \int_{b}^{\infty} \diff r \;
  \epsilon_{\nu}(r) r \sqrt{r^{2} - b^{2}}\nonumber.
\end{eqnarray}

\subsection{The simulated observations}

The integrated spectrum corresponding to each annulus is then redshifted in
terms of both magnitude (as per the appropriate luminosity distance) and
energy. The simulations reported here are for a redshift of 0.017,
corresponding to a luminosity distance of 100\unit{Mpc}. These data are
then converted to \xspec{} table-model format \fits{} files \citep{arna99}.
For added verisimilitude we include the effects of absorption by
multiplying our model spectra with the \xspec{} \phabs{} model, using a
column density $\mysub{N}{H} = 10^{21}$\pcmsq. Since we are now interested
in spatial effects, for response matrices we have used the \chandra{}
\aciss{} data, specifically the AO-2 proposal planning \rmf{} and \arf{}
files\footnote{\texttt{http://asc.harvard.edu/caldb/Aeff/}} for the detector
aim-point position.

The simulated integration time was typically 25\unit{ks}. Noise was added
via counting statistics, but no background file was included. The resulting
fake data were grouped to a minimum of 20 counts per channel to ensure
applicability of the $\chi^{2}$ statistic.

\section{A Chemically Inhomogeneous ICM}

\begin{figure*}
  \begin{center}
    \includegraphics[draft=false,angle=-90,width=2.0\columnwidth]
    {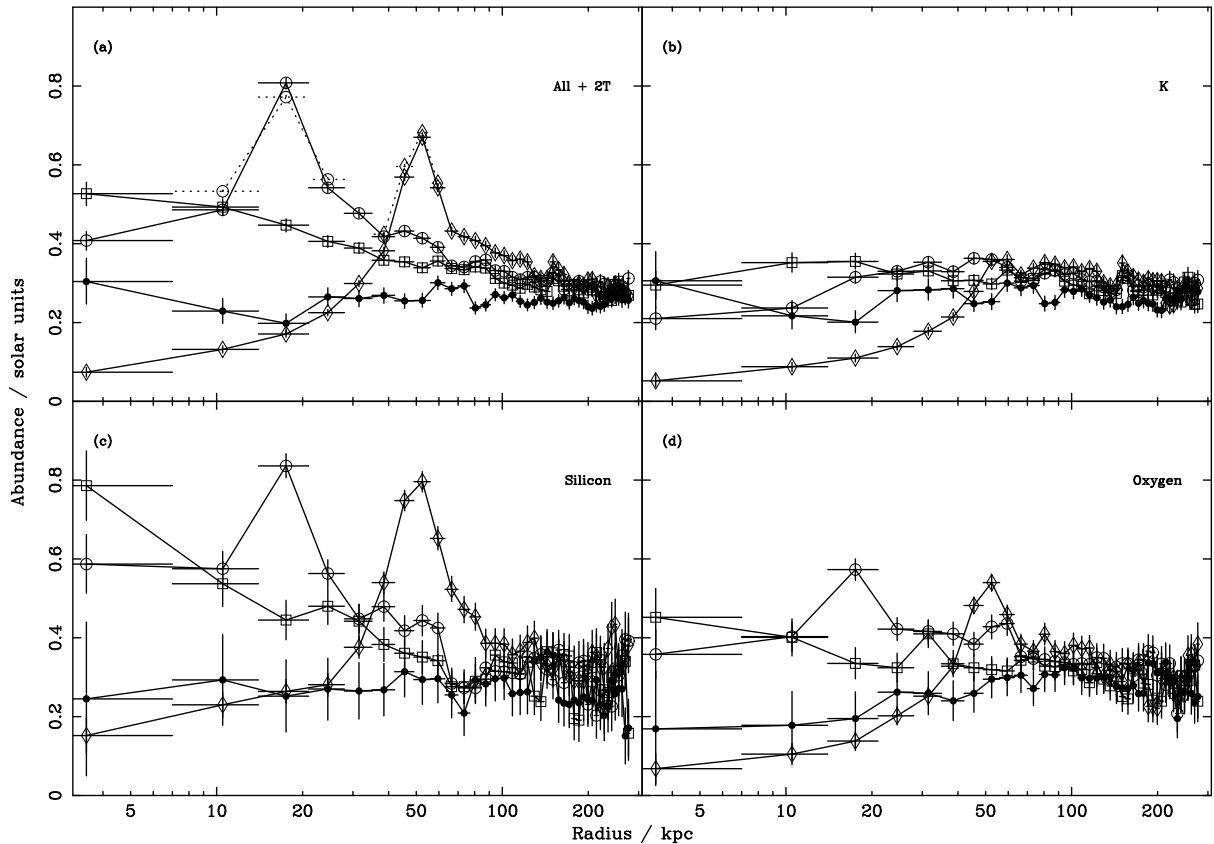}
  \end{center}
  \caption{
    Time evolution of the observed abundance profile for a bimodal
    metallicity. Vertical error bars are 1$\sigma$. See text for model
    properties. Filled circles -- 0.0\unit{Gyr}; open squares --
    1.0\unit{Gyr}; open circles -- 1.2\unit{Gyr}; open diamonds --
    1.6\unit{Gyr}. Unless otherwise stated, all results are for single
    temperature models fitted to the 0.3--7.0\unit{keV} spectral range. The
    simulated observation time was 25\unit{ks} for the upper two panels,
    and 50\unit{ks} for the lower two. Panel \textbf{(a)}: solid lines --
    single temperature \xmekal{} fits; dotted lines -- results of two
    temperature fits, shown where the reduced $\chi^{2}$ for the single
    temperature fits exceeded 2.0. Panel \textbf{(b)}: \xmekal{} fits to
    the 3.0--7.0\unit{keV} (i.e.\ iron K) spectral region. Panels
    \textbf{(c)} and \textbf{(d)}: \xvmekal{} fits for the silicon and
    oxygen abundance, respectively.}
  \label{fig:abund}
\end{figure*}

\subsection{A bimodal metallicity distribution}

Our aim is to investigate the effects of a `clumpy' metal distribution in
the ICM\@. In our model, each radial zone corresponding to the difference
equations may have a distinct set of element abundances. For this work, we
have chosen one of the simplest possible inhomogeneities: 9 out of every 10
zones are taken to be pure H, He in the solar abundance ratio; whilst every
tenth zone has a three times solar abundance of the 13 heavy elements
included by \mekal{} (with He remaining fixed at the solar value). A
spatially varying distribution of this simple form appears identical to a
homogeneous plasma of abundance a little less than one third solar.

Here we present the results for one particular cluster, with parameters as
follows: virial temperature 8.6\unit{keV}{} ($10^{8}$\unit{K}), which from
Eqn.~\ref{eqn:evra} corresponds to a virial radius of 3.4\unit{Mpc}{} and a
halo mass of $2.3\times 10^{15}\Msun$. The radial range was modelled with
5000 zones, corresponding to a resolution of 700\unit{pc}, or around 1.5
arcsec for the redshift in question. The gas fraction was 0.17
\citep[e.g.,][]{etto99}. With a concentration of 5, the NFW scale radius is
700\unit{kpc}, and the associated (singular isothermal sphere) velocity
dispersion is 1300\kmps.

Fig.~\ref{fig:abund} illustrates the results of various \xspec{} fits to the
simulated observations produced for this system. We display only the
central regions where cooling has progressed significantly in the elapsed
time. In each case the specified \xspec{} model was multiplied with a
\phabs{} component. The redshift $z$ and column density \mysub{N}{H} were
fixed at the correct values.

Panel (a) illustrates the results for single temperature \xmekal{} fits to
the spectral range 0.3--7\unit{keV}, in which the plasma temperature,
abundance and normalization were allowed to vary as free parameters.

\begin{figure}
  \begin{center}
    \includegraphics[draft=false,angle=-90,width=0.9\columnwidth]
    {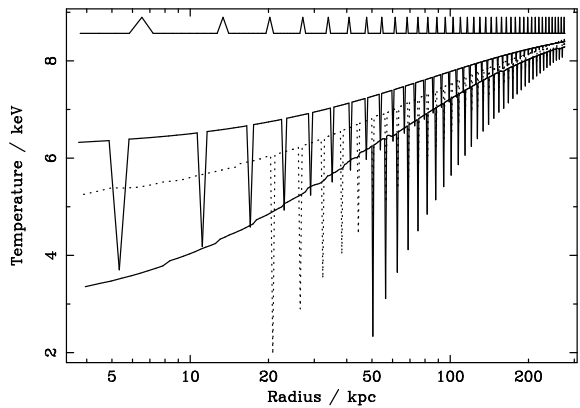}
  \end{center}
  \caption{
    The physical temperature profiles for the model of
    Fig.~\ref{fig:abund}. In order of decreasing average temperature, the
    curves are: 0.0, 1.0, 1.2 (dotted), 1.6\unit{Gyr}.
  }
  \label{fig:truetemp}
\end{figure}

Firstly, note that the initial fitted abundance profile is flat (within the
given error bars), with no high metallicity spikes. This confirms the basic
premise that a varying abundance plasma can resemble a homogeneous gas.
Fig.~\ref{fig:truetemp} shows the evolution of the true (not fitted)
temperature profile of the gas with time. Note that in the initial
conditions the high metallicity zones are at higher temperatures. This is a
consequence of their higher molecular weight (0.63 as opposed to 0.61 in
this specific case). Recall that we require a smooth pressure profile, and
that this is a function of $T/\mu$. These initially high temperatures are
of no consequence for the subsequent evolution of the system.

As the system evolves, it is obviously apparent that the metal-rich regions
cool more swiftly than the metal-poor regions, leading to the rapid
inversion of the `spikes' in the temperature profile shown in
Fig.~\ref{fig:truetemp}. This is an inevitable consequence of the shorter
cooling time of the high abundance gas (Fig.~\ref{fig:tcool}), which in
turn is due to its enhanced radiation (Fig.~\ref{fig:lambda}). The cooling
is naturally most extreme in the central regions where the gas density is
highest, and the two-body bremsstrahlung radiation is most intense. The
disappearance of the innermost spikes in the temperature profile at late
times is due to the removal of the metal-rich gas from the simulation as it
cools below $10^{5}$\unit{K}.

This differential cooling has marked effects upon the measured abundance
profile (Fig.~\ref{fig:abund}a). As time passes, there is a gradual
increase in the derived abundances, particularly in the central regions
(for example, after 1.0\unit{Gyr} the measured value for the central
abundance has risen to 0.5\Zsun). With further evolution, the central
abundance begins to decrease again, leading to the situation where there is
a peak in the abundance profile at an off-centre position. This peak
declines in magnitude and moves outward with subsequent evolution.

This behaviour may be explained as follows. In our simulations, the
metal-rich gas (contributing the emission lines) cools relatively quickly
and enters the regime where line emission is more important, whereas the
metal-poor gas (contributing the bulk of the continuum) cools much more
slowly. The combined consequence of these two processes is an increase in
the strength of the lines relative to the continuum, that is, an increase
in the equivalent width of the lines. It is from the equivalent width that
the fitting procedure obtains the plasma abundance. One could view this as
a form of emission-weighting -- as the metal-rich regions cool the
intensity of their emission increases, so they increasingly dominate the
spectral fits. The decrease in the central abundance at late times occurs
as metal-rich gas there cools out of the X-ray regime and is lost from
sight.

The maximum amplitude of the abundance peak is obviously related to the
underlying spread in metallicity, but it is not obvious how one might use
the former in practice to reconstruct the latter. From the theoretical
point of view there is a lower limit to the extent of the variations that
must exist if they are to be of any use in resolving the cooling flow
problem (i.e.\ the lack of low-energy spectral lines). The upper limit to
the spread depends on what one considers reasonable for the injection
process to produce, assuming there is no active segregation of
metals. Cooling time considerations also become important with extreme
abundance values, if one does not wish the entire central region to become
devoid of metals in the high temperature phases.

Fig.~\ref{fig:abund}a also displays the results of two-temperature
\xmekal{} fits to the data (dotted lines). These are shown only in those
cases where the reduced $\chi^{2}$ for the one-temperature fits was greater
than 2.0. It was necessary to restrict the abundances of the two \xmekal{}
components to be equal, there not being enough information in the spectra
to obtain meaningful constraints on two individually varying abundances.
Even taking this step, the values obtained for the two temperatures were
more often than not very poor limits. The resulting abundances, however,
agree very well with the one-temperature results.

To check that these effects are a genuine consequence of the non-uniform
metal distribution, we have carried out identical simulations where all the
gas has a uniform abundance of 0.3\Zsun. In this case, all four abundance
profiles analogous to those plotted in Fig.~\ref{fig:abund} remain
resolutely flat within the errors at all times as the system evolves.

The results for lower temperature clusters (3 and 5\unit{keV}) are
qualitatively similar. Obviously, with lower temperature systems there is a
greater need for two-temperature models to fit the correspondingly richer
spectra. Also, while for the 8\unit{keV} cluster the maximum height of the
abundance peak (0.9\Zsun) is some three times the average abundance, for
the 3\unit{keV} system the peak has an amplitude roughly double the average
abundance. There is less opportunity for differential cooling if the bulk
of the gas begins life at lower temperatures. The evolution of lower
temperature systems also takes place more rapidly (recall that from
Eqn.~\ref{eqn:tcool} that $\mysub{t}{cool} \sim T^{\frac{1}{2}}$ for
virially-scaled clusters). The general trends of behaviour are, however,
the same. The small-scale metallicity variation scenario would therefore
predict that the off-centre abundance peaks would be proportionately
stronger in higher temperature systems.

\subsubsection{Matters of resolution}\label{subsec:reso}

The size of the spectral annuli is controlled by two factors. Firstly, it
is of course necessary to ensure that they are large enough to collect
sufficient photons in a reasonable integration time to produce a meaningful
signal-to-noise level. Secondly, there is also an issue of numerical
resolution. The size of the annuli is such that each encompasses ten radial
zones from inner to outer edge (although of course owing to projection
effects each annulus receives emission from every zone whose radius is
greater than that of the inner edge of the annulus). This is the minimum
number necessary to avoid artificial oscillations in the fitted abundance
profile, given that the fraction of metal-rich zones is 10 per cent. With a
fraction \mysub{f}{hi} of zones being metal rich, spectral annuli must
encompass at least ${\mysub{f}{hi}}^{-1}$ radial zones.

Alternatively, if the spectral annuli were any smaller then we would be
able to \textit{resolve} the discrete nature of the metal distribution.
Given that each annulus spans a projected radius of 7\unit{kpc}, whilst the
size of the metal rich zones is 0.7\unit{kpc}, it might seem that we are
claiming it would be possible to resolve the inhomogeneities with a
resolution length greater than their size. This is not the case, however,
because the spatial extent of the metal-\textit{poor} zones is nine times
that of the metal-rich zones. When the resolution length is less than or
equal to the greater of those extents, the distribution can in principle be
resolved.

As is standard practice with numerical simulations, we have doubled the
numerical resolution to check that the results are unaffected, and indeed
they are not. This is important in this case for another reason. As well as
confirming the numerical result, it illustrates just how difficult it will
be to probe such metallicity distributions in real-world observations.
Looked at another way, then going from the high resolution simulation to
the default resolution simulation doubles the physical size scale of the
metallicity variations, so that the largest relevant length scale (that of
the metal-poor regions in our model) is just less than the resolution
length of the spectral observations. Yet it does not do anything towards
hinting that there are in fact extreme enrichment variations on just
slightly smaller scales. In other words, detection of such metal variations
will be an all-or-nothing affair -- without adequate spatial resolution,
there is no hope.

Recall as well that this is for a highly simplistic, regular,
one-dimensional distribution. Given an irregular, two- or three-
dimensional distribution pattern, with metal-rich `clouds' drifting in and
out of various lines of sight, then the problem becomes even more
difficult.

If we reduce the size of the spectral annuli below the level we have used,
then we do not immediately begin to resolve the correct (i.e\
0.0--3.0\Zsun) metal distribution. Instead, one begins to detect only
slight variations around the mean level. These will not be significant
unless one has a good quality spectrum with adequate signal-to-noise. The
smaller one makes the spatial resolution element (in order to probe finer
and finer regions), the larger the collecting area required in order to
obtain good-quality spectra in reasonable time-scales. Good spectral
resolution will also be necessary to separate the effects of the varying
enrichments and allow one to constrain two-temperature models with freely
varying abundances for each component. This will depend on just how extreme
the metal variations are. Given that this is all for a regular,
one-dimensional system, with no background file added, one can begin to
appreciate the difficulties that genuine observations present in this
regard.

Accepting that \textit{direct} detection of small-scale ICM metallicity
deviations will be so difficult, we are forced to consider what forms of
\textit{circumstantial} evidence it may prove possible to employ.

\subsubsection{Abundance gradients}\label{sec:grads}

Fig.~\ref{fig:abund}a illustrates that a bimodal abundance distribution
within a cooling flow scenario has the ability to produce an
\textit{apparent} metallicity gradient where one does not in reality exist
(the real mean abundance in our simulations remains unchanged, at least
until metal-rich gas begins to drop out in the inner regions). For example,
consider the results after 1.0\unit{Gyr} of evolution, when the central
abundance has risen to around 0.5\Zsun. At later times the gradient effect
would be even more pronounced if one were not able to resolve the central
abundance drop.

Large-scale abundance gradients were detected in several clusters with
\asca{} and \rosat, for example: Centaurus \citep{fuka94,alle94}; Virgo
\citep{mats96}; AWM~7 \citep{ezaw97}; A496 \citep{dupk00}. The presence of
an abundance gradient appears to be correlated with the existence of a
cooling flow \citep[e.g.,][]{gran01}, although this may not be a true
correlation (cooling flows generate gradients), but merely due to the
mergers which disrupt cooling flows also erasing any abundance gradient.
Here we have demonstrated one mechanism by which cooling flows may actually
give rise to the appearance of such abundance gradients. The interplay
between cooling flows and abundance gradients has been studied previously
by several authors. For example, \citet{alle98} argued that it was the
presence of abundance gradients in cooling flow systems that leads to the
higher emission-weighted metallicities in these sources as compared to
non-cooling flow clusters. \citet{reis96} investigated the ability of
cooling flows to create metallicity gradients by transport of metals.

Recently, \chandra{} studies have confirmed the presence of large-scale
abundance gradients in many clusters. The unprecedented spatial resolution
of \chandra{} has revealed more detail in the profiles. In several cases,
clusters are found to exhibit `peaked' abundance profiles (i.e.\ a positive
radial abundance gradient in the innermost regions, coupled with a negative
gradient further out), very similar to those produced in our simulations.
For example: Centaurus \citep{sand02}, where the negative abundance
gradient peaks at 1.3--1.8\Zsun{} at a radius $\sim 15\unit{kpc}$ before
falling back to 0.4\Zsun{} at the centre; A2199 \citep{john02}, where the
metallicity rises from $\sim 0.3\Zsun{}$ at 200\unit{kpc} to $\sim
0.7\Zsun{}$ at 30\unit{kpc} before dropping back to 0.3\Zsun{} within the
central 5\unit{kpc}; and possibly Perseus \citep{schm02}, where there may
be a high metallicity ring of around 0.6\Zsun at a radius of 60\unit{kpc}.
It is of course intriguing when a hypothesis designed to answer one issue
(lack of low temperature line emission) ends up providing a potential
explanation for another (abundance gradients with off-centre peaks); though
of course there is no shortage of alternative explanations for both these
phenomena.

Extrapolating our results for several Gyr, one would predict that over not
too long a timescale, the central region would become devoid of metals,
with a ring of high metallicity at large (hence easily observable) radii
that would have been detected before the \chandra{} era. This is clearly
unphysical, and is easily explained away by our simple treatment of a
complex problem. We start with fully formed metallicity variations rather
than allowing them to develop; the variation is extreme -- rather than
allowing for several `phases' of differing abundances we use only two, one
with no metals and one with a high abundance; and we do not account for
replenishment of metals through continued enrichment, which would offset
the extremes of behaviour that our simple models predict.

The metallicity distribution in Perseus appears to show no correlation with
the galaxies. Intriguingly, \citet{schm02} also report what may be the
first signs of small-scale metallicity variations in the Perseus ICM,
although the scale of the effect is at the limit of detectability with
current exposure times and is not statistically significant.

A similar off-centre abundance peak is seen in the \chandra{} observation
of the merging cluster A3266 \citep{henr02}. These authors suggest an
alternative explanation for this feature involving the merging subcluster
depositing enriched gas close to the centre of the cluster. Shock heating
is then called upon to preferentially devoid the enriched region of the
lightest, most mobile elements, H and He, so that the metallicity may be
increased without an associated raising of the density, for which there is
no observational evidence. This explanation requires efficient motion of
ions, something which it is not clear can happen even if conduction is
relatively unimpeded. Moreover, these efficient transport properties are
required precisely in the regions where subcluster merging is taking place.
Cold-fronts (see Section~\ref{sec:disc}) are plausible evidence that
transport properties may be highly suppressed in such volumes, presumably
due to separate magnetic structures existing in the merging components. It
is intriguing that off-centre abundance peaks appear to be found both in
clusters with and without cooling flows. If we are looking at the same
phenomenon in both cases (this of course by no means clear), and if a
common mechanism is responsible, this would suggest that it is something
associated with neither the cooling flow process, nor the merger process.

\subsubsection{Equivalent width effects}\label{sec:kgradient}

\begin{figure}
  \begin{center}
    \includegraphics[draft=false,angle=0,width=0.9\columnwidth]
    {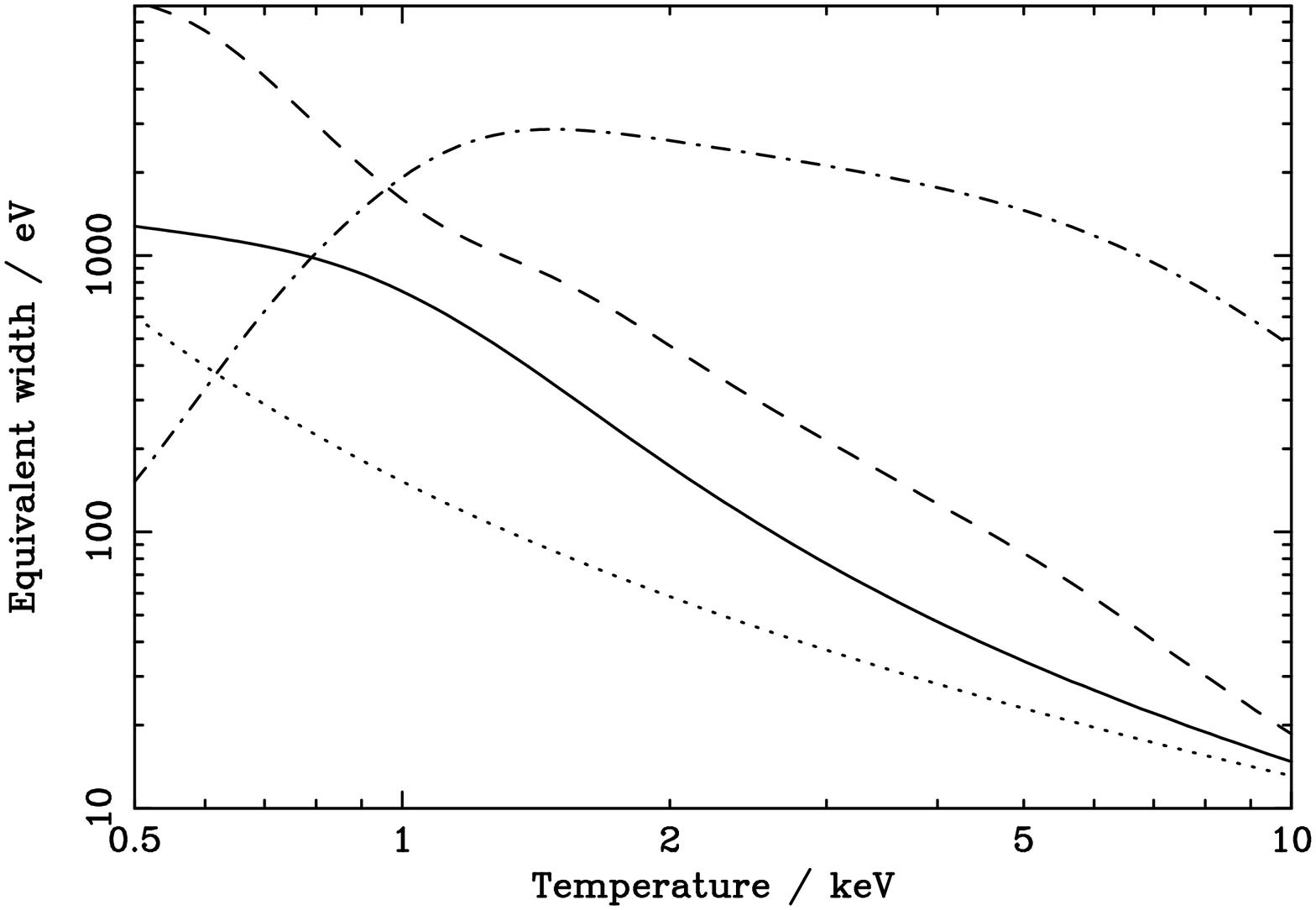}
  \end{center}
  \caption{
    The temperature dependence of the equivalent widths for various ICM
    plasma spectral lines, for a solar metallicity plasma. Dot-dashed iron
    K; dashed iron L; dotted oxygen K; solid silicon K. See text for
    details.
  }
  \label{fig:eqwidths}
\end{figure}

The original \asca{} abundance gradients were clearly detected in the
equivalent width of the iron K line (e.g.\ in Centaurus where
\citealt{fuka94} report an increase by a factor of three on moving to the
central regions of the cluster). Excluding the iron L complex from the
\chandra{} data for Centaurus by simply fitting to the high-energy end of
the spectrum (3--7\unit{keV}), whilst obviously increasing the noise, does
not affect the qualitative nature of the conclusions regarding the
abundance profile \citep{sand02}. Fitting our model spectra just using the
data in this region produces different results, however, as illustrated in
Fig.~\ref{fig:abund}b. The noise level has increased somewhat, but the
qualitative behaviour has also altered. There is still something of an
initial increase in the central abundance at early times, leading once
again to the generation of an apparent gradient. At later times, however,
the highly peaked abundance profile that was present when fitting to the
entire spectrum does not appear. Instead the central abundance merely dies
away. Excluding the iron K lines, however, by fitting only to the spectra
between 0.3--5\unit{keV}{} produces much the same results as using the
entire spectrum.

Thus there is a qualitative difference in behaviour when fitting to the
iron K lines as compared to the iron L lines that must be explained. In
Fig.~\ref{fig:eqwidths} we illustrate the temperature dependence of the
equivalent width for various important ICM spectral lines, calculated from
\mekal{} spectra.

Consider just the results for the iron. Upon cooling from high
temperatures, the equivalent width of the iron K lines increases somewhat,
then rapidly dies away below 1\unit{keV}. The width of the L complex, on the
other hand, increases strongly and monotonically as the temperature
decreases. Consequently, with a bimodal metallicity, the strength of the
iron K lines relative to the continuum increases to a relatively small
extent on cooling, producing the initial slight increase in the central
abundance, before dying away as cooling progresses. The strong monotonic
increase in the iron L width with cooling produces a much greater change
relative to the continuum.

To illustrate this point further, the equivalent widths for the silicon and
oxygen K lines are also plotted in Fig.~\ref{fig:eqwidths}. Both increase
monotonically as the temperature is reduced, but the rate of change for
oxygen is relatively small, whereas silicon shows very similar behaviour to
iron L. The arguments of the previous paragraph are borne out by the
results of \xvmekal{} fits to the model spectra (Figs.~\ref{fig:abund}c and
d show the fits for silicon and oxygen respectively). The qualitative
behaviour of both profiles is the same, but the oxygen profile shows a less
pronounced peak.

\begin{figure}
  \begin{center}
    \includegraphics[draft=false,angle=0,width=0.9\columnwidth]
    {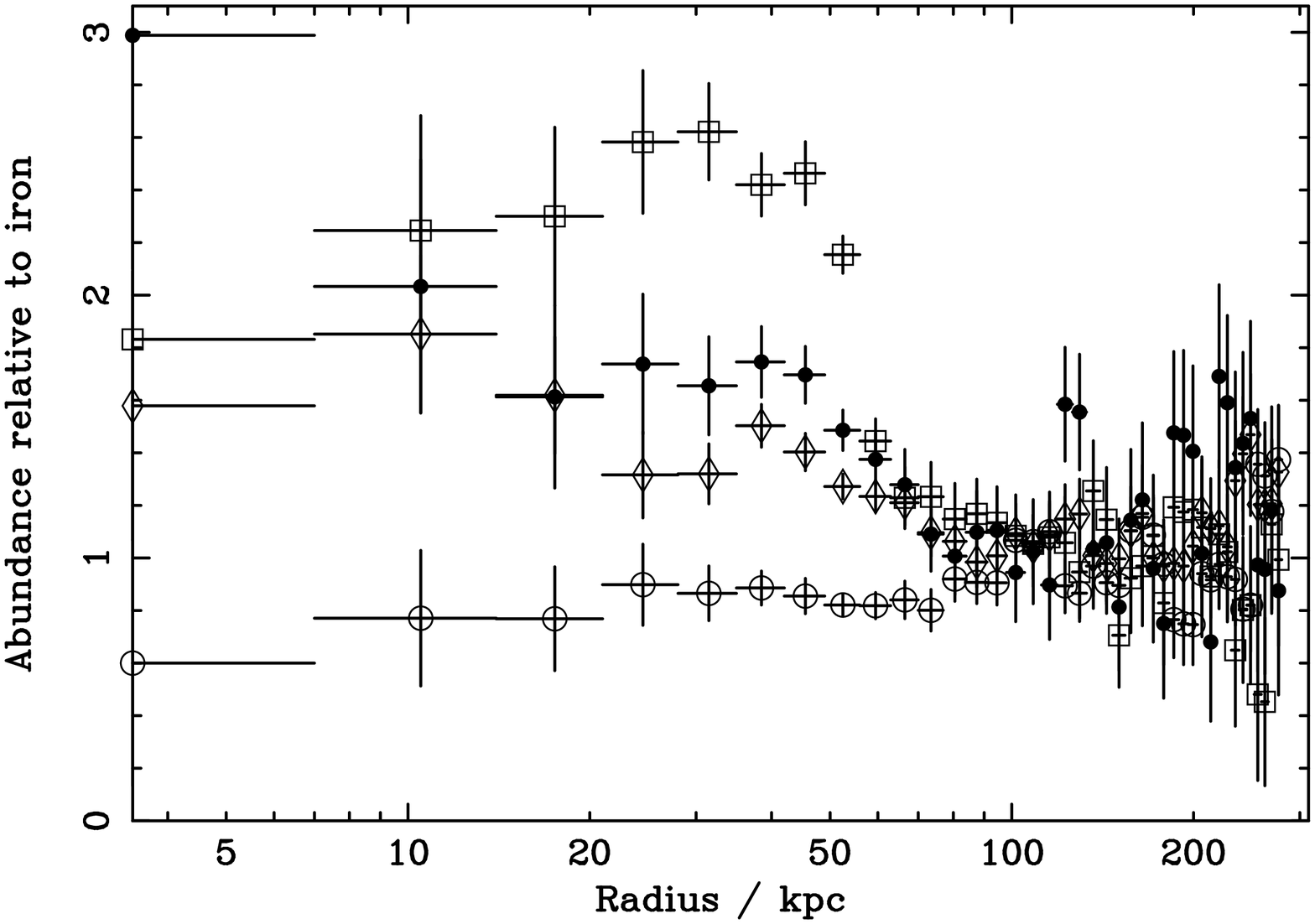}
  \end{center}
  \caption{
    Ratios of element abundance to that of iron. Results are for the last
    of the times plotted in Fig.~\ref{fig:abund}. Filled circles --
    magnesium; open squares -- neon; open circles -- oxygen; open diamonds
    -- silicon.}
  \label{fig:abratio}
\end{figure}

In Fig.~\ref{fig:abratio} we plot the ratio of element abundance to iron
abundance (both in solar units), for various elements, for the last of the
times illustrated in Fig.~\ref{fig:abund}. Neon and magnesium (and to a
lesser extent silicon) appear overabundant, whilst oxygen appears slightly
underabundant.

\subsection{Genuine abundance gradients}

\begin{figure*}
  \begin{center}
    \includegraphics[draft=false,angle=-90,width=2.0\columnwidth]
    {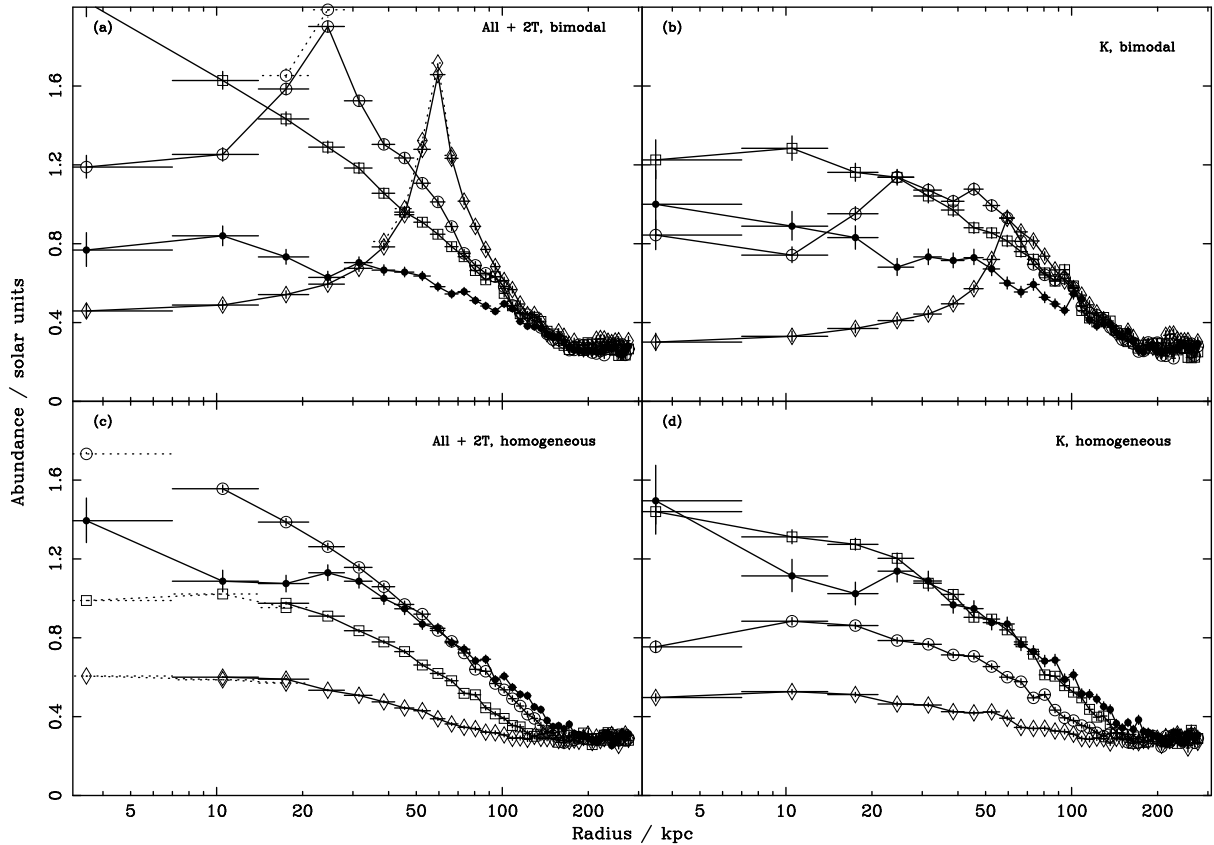}
  \end{center}
  \caption{
    Time evolution of the observed abundance profile for clusters with
    genuine abundance gradients. Vertical error bars are 1$\sigma$,
    simulated observation time 25\unit{ks}, all results for single
    temperature \xmekal{} models unless otherwise stated. Top two panels,
    bimodal metal distribution: filled circles -- 0.0\unit{Gyr}; open
    squares -- 0.4; open circles -- 0.5; open diamonds -- 0.8. Bottom two
    panels, homogeneous metal distribution: filled circles --
    0.0\unit{Gyr}; open squares -- 1.3; open circles -- 2.5; open diamonds
    -- 3.8. Left-hand panels: fits to 0.3--7.0\unit{keV} region, dotted
    lines show results from two temperature fits where the reduced
    $\chi^{2}$ for the single temperature fits exceeded 2.0. Right-hand
    panels: fits to the 3.0--7.0\unit{keV} (i.e.\ iron K) spectral region.
    }
  \label{fig:gradabund}
\end{figure*}

Given that flat metallicity profiles coupled with bimodal distributions are
unable to fully reproduce the observed abundance trends, we have considered
also clusters with genuine gradients in their abundance profiles. We have
parameterized the profile in the following manner:
\begin{equation}\label{eqn:zprof}
  Z(r) = \left\{
    \begin{array}{ll}
      \bar{Z} + (\mysub{Z}{c} - \bar{Z})
      \left( 1 - \frac{r}{r_{0}}\right)^{n} &
      r < r_{0} \qquad (n \ge 0)\\
      \bar{Z} & r \ge r_{0}
    \end{array} \right.
\end{equation}
where $\bar{Z} = 0.3$ is the mean cluster abundance; $\mysub{Z}{c} = 2.0$ is
the central abundance; $r_{0} = 200\unit{kpc}$ is the extent of the
gradient; and $n = 1.5$ controls its severity.

We have carried out simulations both with homogeneous metallicities (i.e.\
the distribution smoothly follows the profile), and with a bimodal
behaviour similar to before (i.e.\ 90 per cent of the gas is metal-poor,
the other 10 per cent has ten times the abundance dictated by
Eqn.~\ref{eqn:zprof}). The results are displayed in
Fig.~\ref{fig:gradabund}. Once again, the bimodal system develops an
off-centre abundance peak, of a larger amplitude than before. This time,
the trend is reproduced when fitting solely to the high-energy part of the
range, panel (b). In the homogeneous system, on the other hand, no such
peak develops, rather the abundance declines reasonably smoothly back down
towards a flatter state. Note that the temporal evolution is in both cases
faster, owing to the higher abundances that are present.

\subsection{Cooling flow equivalent widths}

We have used simulations similar to those described in the previous section
to produce synthetic \xmm{} \rgs{} spectra in order to investigate in more
detail the ideas of Section~\ref{sec:cf_spec} on the suppression of the
low-temperature lines. These broadly reproduce the same results as the
isobaric model, Eqn.~\ref{eqn:cflow}. They do reveal another factor,
however, which conspires to reduce the line suppression. When geometrical
(i.e.\ position within the cluster potential) considerations are taken into
account, the metal-rich gas always has a larger cooling radius at any
instant than the metal-poor gas, as a fairly obvious consequence of its
faster cooling rate. In the case of an 8.6\unit{keV} cluster, for example,
where 10 per cent of the gas is made to be metal-rich, the fraction of
cooled mass originating from the enriched phase asymptotes to about 1/8,
rather than 1/10. This reduces still further the equivalent width
suppression we can expect from a bimodal distribution of metals.

\section{Discussion}\label{sec:disc}

To address the question of how small-scale ICM abundance variations might
come about, supernovae can be considered as essentially point sources (on
ICM scales) of extremely high metallicity gas. In of themselves, these have
structures that are far from simple \citep[e.g.,][]{hugh00}. One way in
which escape of the enriched gas from the potential well of the host galaxy
into the ICM may occur is via superwinds \citep[e.g.,][]{heck01} (although
other processes such as ram-pressure stripping and pregalactic winds are
doubtless significant). Superwinds have a complex, multiphase structure in
which it is difficult to probe directly the energetic, enriched gas driving
the wind. There is, however, clear evidence for complicated structure in
the wind on very small scales in the optical and soft X-ray
\citep[e.g.,][]{stri01}. In our opinion, the question ought to be posed the
other way around: how might the ICM become uniformly enriched?

There are essentially two issues that determine whether or not the
situation as we have chosen to model it is physically realistic: i) how
are the metals injected into the ICM, and ii) once in the ICM, do the
metals move to spread out over a wide area or remain confined? The answers
to both these questions remain uncertain at present. To address the second
question first, following \citet{spit62}, we may express the deflection
time (average time for a cumulative deflection of $90^{\circ}$) for
particles of mass $m$, charge $Ze$ diffusing through field particles of
mass $\mysub{m}{f}$, charge $\mysub{Z}{f}e$, density $\mysub{n}{f}$ as
\begin{equation}
  \tau = \frac{6 \sqrt{3} \pi \epsilon_{0}^{2} \mysub{k}{B}^{3/2}}{e^{4}}
  \frac{m^{1/2} T^{3/2}}{Z^{2} \mysub{Z}{f}^{2} \mysub{n}{f}
    \mathcal{F}(x) \ln{\Lambda}}
\end{equation}
where
\begin{eqnarray}
  x & = & \sqrt{\frac{3 \mysub{m}{f}} {2 m}}\\
  \mathcal{F}(x) & \equiv & \erf{x} + \frac{\diff}{\diff\,x}
  \left(
    \frac{\erf{x}}{2 x}
  \right)
\end{eqnarray}
with $\erf$ the standard error function, $\ln{\Lambda}$ the Coulomb
logarithm, and assuming that all species are in thermal equilibrium at
temperature $T$. The factor $\mathcal{F} \le 1$ is the Chandrasekhar
correction for the finite mass of the field particles.

Numerically
\begin{equation}
  \tau \approx 0.26 \unit{Myr}\quad A^{1/2}
  \left( \frac{T}{10^{8}\unit{K}} \right)^{3/2}
  \left( \frac{\mysub{n}{f}}{\unit{cm}^{-3}} \right)^{-1}
  \frac{1}{Z^{2} \mysub{Z}{f}^{2} \mathcal{F} \ln{\Lambda}}
\end{equation}
with $A$ the test particle mass in units of the proton mass.

As was shown by \citet{reph78} for the case of sedimentation, in the ICM
the contributions from field particles other than protons (particularly
helium nuclei) are significant. Summing over the elements in a $0.3\Zsun$
plasma with iron nuclei as test particles, we find that $\tau$ is reduced
by around a factor of two from the value for a pure proton plasma. At
$10^{8}\unit{K}$, and with $\mysub{n}{H} = 10^{-3}\unit{cm}^{-3}$, we find
the corrected deflection times for iron and helium nuclei are
$\mysub{\tau}{Fe} \approx 0.3 \unit{Myr}$, $\mysub{\tau}{He} \approx 4
\unit{Myr}$.

From standard kinetic theory, the root-mean-square three-dimensional
distance a particle will diffuse in a time t is given by
\begin{equation}
  \mysub{r}{rms} \; = \; \sqrt{2 \lambda \mysub{v}{rms} t}
  \; = \; \mysub{v}{rms} \sqrt{2 \tau t}
  \; = \; \sqrt{\frac{6 \mysub{k}{B} T \tau t}{m}},
\end{equation}
where $\lambda = \tau \mysub{v}{rms}$ is the mean free path. Hence we can
estimate that iron nuclei may diffuse about 20\unit{kpc} in 1\unit{Gyr}. To
maintain small-scale abundance variations over significant timescales in
the ICM, we therefore require a strong suppression of diffusion. This
result is to be expected, as it is equivalent to the strong suppression of
conduction that has traditionally been invoked for the multiphase cooling
flow picture. Magnetic fields have normally been appealed to as the
causative agent to dampen transport properties in an ionized gas. Early
calculations \citep[e.g.,][]{trib89} appeared to show that a tangled
magnetic field would indeed produce a strong reduction in transport. More
recent calculations considering a chaotic field with turbulence extending
over a wide range of length scales indicate that the suppression may in
fact be minimal \citep{nara01}.

In practice, simple diffusion is probably unlikely to be the limiting
factor for the spread of ICM metals. Convection, turbulence due to radio
sources, galactic wakes, etc., will also play roles to varying degrees. If,
however, the abundance drops observed with \chandra{} in the central
regions of the ICM for several clusters are genuine, then this will imply
limits on the amount of convection or mixing that can have taken place
(else these features would have been smoothed out). It does not seem
unreasonable to explore some of the consequences of small-scale metallicity
variations, and to keep the possibility in mind during spectral analyses.

The question as to whether or not the metals in the intracluster and
intergalactic media are homogeneously distributed remains an open one.
There is a non-negligible scatter in Galactic stellar metallicities for
stars of all ages (the scatter increases for low metallicity stars), which
has been taken to suggest that that Galactic disc has been chemically
inhomogeneous throughout its development \citep[e.g.,][]{mcwi97}. Classical
galactic chemical evolution models have tended to assume instantaneous
dispersion of synthesized elements.

The average metallicity of the intracluster medium on large scales has for
some time been reasonably well established at roughly 1/3 solar in most
cases, both for nearby \citep{edge91a} and distant \citep{mush97} clusters.
Until comparatively recently, however, there has been little information on
how the ICM metals might be distributed on finer scales. There is, in our
opinion, no particular reason why the whole ICM should be uniformly
enriched to the same metallicity, although this might be one's natural
assumption. As we discuss in Section~\ref{subsec:reso}, direct detection of
small-scale abundance variations in the ICM will be very difficult.
Consequently, we feel it is impossible to rule out such variations at the
present time.

Recently, a deal of support has been given to the idea that thermal
conduction might be operating in cluster cores at significant levels
\citep[e.g.,][]{nara01,voig02,fabi02}. Whilst this runs contrary to the
established multiphase cooling flow picture \citep{nuls86}, it seems to
have a degree of success in matching the observed temperature profiles. It
is certainly the case, though, that at least in some regions of the ICM,
conduction is highly suppressed. This is revealed by the `cold-fronts' seen
in several cluster cores, e.g.\ A2142 \citep{mark00}. It was shown by
\citet{etto00} that such features, which correspond to a sharp jump in the
surface brightness profile, indicate a strong (factors of several hundred)
\textit{local} suppression of the thermal conductivity, else the
temperature discontinuity that is responsible for the brightness jump would
be quickly washed out. If thermal conduction is suppressed, then ion
movement must be even more restricted, since net heat conduction may still
take place without individual electrons travelling a significant distance.
If thermal conduction is operating with a degree of efficiency, however,
then the same need not necessarily apply to ion motion. If this were the
case, though, there would be the possibility of sedimentation of the heavy
elements \citep{fabi77,reph78,fabi02}. Cold-fronts of course only provide
direct evidence for inhibited thermal conduction across the fronts
themselves. It is possible (though as yet far from proven) that conduction
may reach the \citet{spit62} value elsewhere \citep[e.g.,][]{voig02}. If
conduction is operating efficiently, it does not necessarily rule out
small-scale metallicity variations, but it does weaken the case, both from
the point of view of reducing/removing the motivation for them (suppressing
the low-temperature cooling flow line emission by suppressing the low
temperature gas), and by making it less likely that such conditions can
persist.

Our modelling is simplistic, and it may be argued that the conditions as we
have chosen to represent them are not physically relevant. For example,
there is no possibility of segregation of elements based on weight (that
is, it is not possible for the metal-rich gas to sink to the centre). This
is a consequence of our adopting the theoretical framework of
\citet{nuls86}, which requires that in a multiphase flow the various phases
co-move, i.e.\ there is a single velocity profile. The arguments for this
idea are set out in detail in \citet{nuls86}. Also, the timescale over
which the changes in the abundance profile occur is somewhat short. This is
a result of the rather unphysical initial condition; starting out with
regions enriched to 3\Zsun{} and not allowing any replacement of cooling
metals. A more realistic treatment would allow the metallicity to build up
initially with time and then allow for some replenishment, and possibly
also for different distributions of SNe Ia and II products. However, we are
not concerned here with producing detailed models of the evolution, but
rather in seeing what the general trends of behaviour might be. Note that
because we have restricted our attention to single-phase models, there is
no real transport of material by the cooling flow, since all the mass loss
(except for the metal-rich zones) takes place in the centre. \citet{reis96}
demonstrated the ability of multiphase cooling flows to create (genuine)
abundance gradients through transport of injected materials. Thus we may
expect that a more sophisticated multiphase treatment of this process might
reveal more complex effects.

Given that the abundance gradients observed with \asca{} and \chandra{} are
seen clearly in the iron K line, the results of Section~\ref{sec:kgradient}
imply that this cannot be solely due to a bimodal distribution of
metallicities. The results are not, however, inconsistent with such a metal
distribution, at least in the simple scenario outlined here. At the very
least, we have presented another reason to be wary of the iron L complex
when fitting to X-ray spectra \citep[e.g.,][]{fino00}. In our case, it is
not due to the complex atomic physics of the L shell (the code we use to
generate the spectra is the same as that we use to fit them), but rather to
the temperature dependence of the equivalent width. Recall that iron is
really the only ICM element for which one has two strong spectral
indicators (K and L); for other elements only the K lines are useful.
Playing devil's advocate, one could therefore imagine a situation in which
the observed gradients in the \textit{iron} profile are genuine, but those
in the other elements are due to a process such as the one outlined here.
This would, for example, severely impact on estimates of the SNe Ia:II
importance ratio. Of course, this simple model would not explain the
matching radial scales of the variations for iron and the other elements in
this case. Nor could it explain any correlation between the iron profile
and the visible light in the central regions of the cluster.

Another process which may give rise to the same sort of radial abundance
profile as those seen here (namely a peak in the abundance at an off-centre
position) is resonant scattering \citep{gilf87}. This acts to redistribute
photons from the central regions of the cluster (where the optical depth is
highest) to a surrounding ring. See the work of \citet{math01} for an
application of these ideas to M87. Computation of detailed optical depths
requires knowledge of the velocity structure of the gas along the line of
sight. We will not comment on the resonant scattering issue here, except to
say that to some extent the ideas of this paper and those of resonant
scattering are in conflict. As was pointed out by \citet{wise92}, clumping
of the ICM reduces the amount of scattering that takes place. X-ray surface
brightness profiles depend on the rms density $\sqrt{\langle n^{2}
\rangle}$, whereas electron scattering depends on the mean density $\langle
n \rangle$. The rms of a set of numbers is always greater than their mean
(unless the numbers are all equal). Thus increasing the degree of clumping
in the plasma reduces the mean density relative to the rms density and so
reduces the relative effect of scattering.

Another possibility is that the `extra' metals in the central regions of
many clusters are to be associated with the central cD galaxy.
\citet{maki01} have suggested an alternative explanation for the enhanced
emission seen at the centre of many clusters, which has traditionally been
attributed to the cooling flow phenomenon by many researchers. Instead,
these authors suggest the excess may be associated with the
inter\textit{stellar} medium of the central cD galaxy. One argument invoked
against the cooling flow interpretation is the fairly frequent presence of
metallicity increases near cluster centres. Our present work suggests a
mechanism by which such effects may indeed be produced by cooling flows.
The reality or otherwise of any central dips in abundance would be an
important discriminant for these two interpretations.

\citet{bohr02} also give some consideration to the possible effects of
inhomogeneous metallicities on the observed abundances (section 2.2 of
their paper). Unlike our consideration of individual emission lines, they
look at the overall observed metallicity of the spectrum, as inferred from
the global ratio of power in all emission lines to power in the continuum.
These authors also discuss some comparisons between the shape of the
spectrum around the 1\unit{keV} region for M87 and the predictions of the
bimodal model. All hypotheses live or die through comparison with data, so
such investigations are highly necessary, but there is a large parameter
space to investigate. And as they point out, such checks must be made on a
source-by-source basis. They find a poor fit between the actual spectral
shape and the predictions of the bimodal hypothesis (e.g.\ their fig.~6).
Note, however, that this is for the case where the normalizations of the
metal-rich and metal-poor phases are `roughly equal'. We would not expect
such a division to be successful in reducing the EW of the low temperature
lines by an appreciable amount.

Our main result is that the observed equivalent width suppression for low
temperature emission lines in a cooling flow spectrum due to a metal
distribution which is inhomogeneous on small-scales is not as great as one
might expect. For example, if all the metals are concentrated in 10\% of
the gas, the suppression of the low-temperature lines relative to those
from high temperature gas is only about a factor of 3
(Section~\ref{sec:cflow_ew}). It seems unlikely, therefore, that this
method in isolation could produce a reduction in equivalent width equal to
that seen in data, without pushing the bimodality of the metal distribution
to extreme levels. There is an effect, but it is not large enough. We have
also shown that small-scale metallicity variations can give rise to
interesting effects in the observed abundance profiles
(Section~\ref{sec:grads}) as compared to the true profiles. Such effects
would give rise to serious difficulties in terms of interpreting abundances
in cluster central regions. The possibility of small-scale metallicity
variations ought to be borne in mind when analysing high resolution X-ray
spectra of cluster central regions.

\section*{Acknowledgments}

We thank PPARC (RGM) and the Royal Society (ACF) for support. RGM is
grateful to Steve Allen and Enrico Ramirez Ruiz for useful suggestions, and
to Robert Schmidt for many helpful discussions.

\bibliographystyle{mnras}

\end{document}